\documentclass{dune}
\pdfoutput=1

\input{common/preamble}
\usepackage{lineno}
\begin{document}

\pagestyle{titlepage}


\pagestyle{titlepage}

\date{}

\title{\scshape\Large Snowmass Neutrino Frontier: \\
NF04 Topical Group Report\\
\normalsize Neutrinos from natural sources
\vskip -10pt
\snowmasstitle
}


\renewcommand\Authfont{\scshape\small}
\renewcommand\Affilfont{\itshape\footnotesize}

\author[1]{Y. Koshio}
\author[2]{G.D. Orebi Gann}
\author[3]{E. O'Sullivan}
\author[4]{I. Tamborra}

\vspace{-0.5cm}
\affil[1]{Okayama University, Okayama, Japan}
\affil[2]{University of California, Berkeley, Berkeley, CA, USA and Lawrence Berkeley National Laboratory, Berkeley, CA, USA}
\affil[3]{Uppsala University, Uppsala, Sweden}
\affil[4]{Niels Bohr Institute, University of Copenhagen, Copenhagen, Denmark}


\maketitle

\renewcommand{\familydefault}{\sfdefault}
\renewcommand{\thepage}{\roman{page}}
\setcounter{page}{0}

\pagestyle{plain} 
\clearpage
\textsf{\tableofcontents}





\renewcommand{\thepage}{\arabic{page}}
\setcounter{page}{1}

\pagestyle{fancy}

\fancyhead{}
\fancyhead[RO]{\textsf{\footnotesize \thepage}}
\fancyhead[LO]{\textsf{\footnotesize \rightmark}}

\fancyfoot{}
\fancyfoot[RO]{\textsf{\footnotesize Snowmass 2021}}
\fancyfoot[LO]{\textsf{\footnotesize NF04 Topical Group Report}}
\fancypagestyle{plain}{}

\renewcommand{\headrule}{\vspace{-4mm}\color[gray]{0.5}{\rule{\headwidth}{0.5pt}}}



\clearpage
\section{Executive Summary}

\subsection{Physics of neutrinos from natural sources}
This report covers a broad range of neutrino sources, from low-energy  neutrinos from the early universe to ultra high-energy sources.  We divide this report by source, and discuss the motivations for pursuing searches in each case, the current state of the field, and the prospects for future theoretical and experimental developments.  
We consider neutrinos produced in the early universe; solar neutrinos; geoneutrinos; supernova neutrinos, including the diffuse supernova neutrino background (DSNB); neutrinos produced in the atmosphere; and high-energy astrophysical neutrinos.

Strong support for this program is essential to preserve and grow our ability to use neutrinos to probe the universe.  These areas of study offer unique insights and opportunities, including the ability to probe the interaction of neutrinos with matter; constraints on the nature and mass of the neutrino that are complementary to, and provide invaluable input to, neutrinoless double beta decay and direct mass searches; an understanding of the formation, evolution, and eventual demise of stars; and flavor physics.

The detection of the cosmic neutrino background will be crucial to unveil the Dirac or Majorana nature of neutrinos, as well as obtain information on their absolute mass. In this endeavor, PTOLEMY will play a crucial role. Upcoming cosmological surveys will be instrumental to probe the early universe physics and infer the eventual existence of new physics. To this purpose, it remains essential to progress in our understanding of neutrino flavor mixing in the early universe.

Future precision measurements of solar neutrinos offer the potential to resolve uncertainties in the metal content of our Sun, to inform our understanding of stellar evolution, to probe matter effects in a unique environment, to search for non-standard interactions and new physics effects, and to perform precision tests of the three-flavor neutrino mixing paradigm.  A suite of next-generation detectors is under development, which leverage novel technology to improve background discrimination, in order to enhance sensitivity to these low-energy neutrinos.  A similar suite of detectors is also sensitive to measure antineutrinos produced within the earth -- so-called geoneutrinos.  New measurements of the geoneutrino flux will inform models for heat production within the Earth.  Of particular interest are measurements of the U/Th ratio; a measurement of the flux produced from the mantle, which can be inferred using a suite of results from different geographical locations, or measured directly with an ocean-bottom detector; and the potential to detector the K antineutrino flux.

The next galactic supernova is expected to be detected in photons, neutrinos, and gravitational waves and will carry a formidable amount of information on the physics of the collapse of massive stars. This multi-messenger detection will be crucial to test our conjectures on the supernova mechanism. To this purpose, progress is being made in order to use neutrinos to alert astronomers and gravitational wave physicists of the core collapse as soon as possible through the SNEWS network. Neutrinos will carry clear signatures of the nature of the stellar collapse, e.g.~in the instance of black hole formation. Despite swift progress in the field in the last decade, the impact of neutrino flavor mixing in the dense supernova core remains to be assessed and it is one of the main goals for the next decade, due to its  implications on the supernova hydrodynamics, nucleosynthesis and detectable neutrino signal. In order to maximize the physics we will extract from this once-in-a-generation event, it is imperative to have neutrino experiments with high uptime and using a variety of detection technologies in order to enhance the statistics of the measurement, but also to ensure measurements of all neutrino flavours. It is expected that the next supernova burst will test the existence of new physics or place stringent constraints on it; however, to this purpose a consistent modeling of new physics in the source is mandatory. The upcoming DSNB detection will push low-energy neutrino astronomy to extra-galactic scales. Various upcoming experiments aim to enable new physics avenues through  the DSNB detection, which will provide complementary insight on the supernova population as well as on new or exotic physics. 

Research into atmospheric neutrinos has had great success in the last two decades, most notably including the first discovery of neutrino oscillations.
In atmospheric neutrino observations by next-generation detectors, planned for operation in the near future, it is expected that the neutrino oscillation parameters, especially $\theta_{23}$ octant, mass ordering, and $\delta_{CP}$, will be determined with high accuracy.
These detectors also offer sensitivity to physics beyond the standard model, such as sterile neutrinos, Lorentz invariance violation, non-standard interactions, and CPT violation, which have not currently been observed.
Precise measurements of the atmospheric neutrino flux are also important topics for non-oscillation physics such as nucleon decay, since the atmospheric neutrino signal forms a dominant background to such searches.
To this end, reductions in the uncertainty of hadron production by primary proton interactions forms an important component of the program.
In addition, there is interest in searches for prompt neutrinos from atmospheric charm production by cosmic rays.

The advent of real-time astronomy through high-energy neutrinos has already led to a handful of likely observations of cosmic sources connected to high energy neutrino production. The expected growing numbers of such detections as well as the larger statistics enabled by upcoming neutrino telescopes will allow to probe particle acceleration in the sources as well as the mechanisms powering cosmic accelerators. In the next decade, we expect the emergence of large neutrino telescopes which will use radio detection to capture neutrinos up to ZeV in energy. This will enable discoveries of cosmic acceleration of neutrinos at the highest energies. In addition, information on the neutrino properties and new physics can be established at these extreme energies for the first time. 

\subsection{Enabling technologies}
Neutrinos from natural sources typically fall into a few energy regimes, and we observe synergies in the experimental techniques and detector requirements within those regimes.

Low-energy neutrinos, in the MeV to 10s of MeV range, have been successfully interrogated using a range of techniques, including radiochemical experiments, water Cherenkov detectors, organic liquid scintillator detectors, noble liquid detectors, and solid state detectors.  Future progress will hinge on critical developments, including fast and spectrally-sensitive photon detection, novel scintillating materials, new purification techniques, and low-background detectors.  These detector developments are closely coupled with improvements in readout electronics that can handle the dynamic range of physics addressed in these detectors, and sophisticated analysis techniques.

At the high energy end of the neutrino spectrum, the main challenge lies with the low flux of signal. The required large instrumented volumes drive the need for powerful yet inexpensive sensors. The field has seen great success with the use of optical sensors deployed in large neutrino telescopes. Looking towards the future, the development of optical sensors that use a collection of "multi-PMTs", smaller photosensors in a single unit that allow for more information about the directionality of the incoming photons, seems promising. In order to instrument the volumes required for the highest energy astrophysical neutrinos beyond the PeV-scale, where the flux is even lower, the development and deployment of inexpensive radio receivers, both in-ice as well as balloon-borne or space-based missions to enhance the observable volume, are required.   

\subsection{Synergies}
All areas within this topical group rely heavily on continued theoretical development, and we wish to emphasise the need for continued, ongoing support in this area.  There is also strong synergy with the computing frontier, as theoretical simulations of astrophysical objects and their related microphysics move towards a new level of complexity and multi-dimensional modeling; in parallel, synergies with the computing frontier will be essential as experiments scale up and data production moves into PBs, along with the need for ever-more sophisticated analysis techniques to enhance sensitivity and fully leverage detector capabilities. As neutrinos play an essential role in multi-messenger astronomy, synergies are also foreseen with the cosmic frontier and with the astronomy community in order to take full advantage of upcoming multi-messenger discoveries.


\clearpage
\cleardoublepage
\section{Early universe neutrinos}

\subsection{Motivation and current understanding}
The Early Universe constitutes an ideal laboratory to investigate fundamental physics as well as the eventual existence of New Physics. 

\subsubsection{Cosmic neutrino background}
The cosmic background of neutrinos (CNB) is a relic of the Universe when it was about $1$~s old~\cite{Vitagliano:2019yzm,ParticleDataGroup:2020ssz,Lesgourgues:2006nd}. It consists of about $112$~cm$^{-3}$ neutrinos and antineutrinos of all flavors. The CNB  has not been detected yet, however ongoing efforts aim to achieve its detection threshold. The CNB detection will  shed light on the Dirac or Majorana nature of neutrinos, the absolute scale of neutrino masses, as well as open promising avenues in the field of dark-matter astronomy. 

\subsubsection{Neutrino properties through cosmological data}
Upcoming cosmological data could be especially relevant to investigate the yet unknown neutrino properties~\cite{ParticleDataGroup:2020ssz}. In fact, while 
we routinely measure  the mass-squared differences of neutrinos from flavor oscillation experiments, the neutrino mass ordering is still uncertain, but it is expected to be measured through neutrino experiments within the next decade. On the other hand, the absolute mass of neutrinos is still unknown. The  cosmological measurement of the latter will complement ongoing searches, e.g.~the ones carried out by the KATRIN Collaboration which  aims to measure the electron endpoint spectrum in tritium beta decay~\cite{KATRIN:2022kkv,Aker:2021gma,KATRIN:2021dfa}. The eventual existence of a tension between cosmological and experimental measurements could provide new insights on particle physics and astrophysics. 

The radiation density in the early universe consists of the contribution of neutrinos and photons, after the decoupling of electrons and positrons~\cite{Lesgourgues:2006nd}. Before  neutrinos become non-relativistic, the radiation density is parametrized in terms of the effective number of  thermally excited neutrino degrees of freedom, $N_{\rm eff}$. The expected value of the effective number of radiation species is $N_{\rm eff} = 3.045$, where the deviation from $3$ takes into account the residual neutrino heating by electron and positron annihilations, in addition to other small corrections. It is important to measure $N_{\rm eff}$ precisely as it could provide indications on the eventual existence of New Physics, such as extra sterile neutrino families or non-standard interactions of neutrinos. Despite the existence of possible ways to evade the $N_{\rm eff}$ constraints for New Physics scenarios,  another challenge concerning sterile neutrinos revolves around the fact that the mass scale of these particles is unknown and we ignore their production mechanism in the early universe, despite the fact that they could have a major impact on the cosmological observables.

\subsection{Future prospects}

\subsubsection{Prospects for theoretical developments}

The measurement of the absolute neutrino mass is one of the main goals of Stage-4 CMB experiments~\cite{AlvesBatista:2021gzc,ParticleDataGroup:2020ssz,Amendola:2016saw,Abazajian:2019tiv} that are expected to pin down the neutrino masses with an uncertainty of up to $15$~meV. Stage-4 CMB experiments also promise to achieve a precision of about $1\%$ in the determination of $N_{\rm eff}$~\cite{Green:2019glg}. The measurement of the primordial helium and deuterium abundances will also be essential and is planned to be achieved through the advent of 30 m-class telescopes~\cite{Cooke:2017cwo}. 

Such upcoming measurements will need to be contrasted with simulations of the evolution of the early universe at the time of neutrino decoupling in order to constrain any New Physics scenarios. However, one of the  theoretical challenges  is to  provide a self-consistent solution of the neutrino quantum kinetic equations including scattering as well as entropy flow. These kinds of calculations are currently not available as they are beyond the state-of-the-art computationally, however they are crucial to disentangle subtle effects due to New Physics from  standard scenarios~\cite{Fuller:2011qy,Hannestad:2012ky,Ruchayskiy:2012si,Gelmini:2019esj,Nygaard:2020sow,Barenboim:2020vrr,Hasegawa:2020ctq}. An important and unsolved issue concerns the modeling of neutrino flavor mixing in the context of the early universe~\cite{Tamborra:2020cul,Capozzi:2022slf,deSalas:2016ztq,Froustey:2020mcq,Hansen:2020vgm,Mangano:2005cc}. In particular, because of the large density of neutrinos, the coherent forward scattering of neutrinos onto each other makes the flavor evolution non-linear, leading to counter-intuitive effects in the flavor evolution. As future cosmological surveys become available, they will enable the measurement of cosmological variables with high precision, hence the exact role played by neutrino mixing needs to be clarified. In addition, dark sector scenarios may mimic the phenomenology of the Standard Model, with subtle implications for the physics of the early universe yet to be grasped.

\subsubsection{PTOLEMY} 
PTOLEMY is a landmark project with the goal of detecting the CNB~\cite{PTOLEMY:2019hkd}. In order to detect neutrinos, PTOLEMY relies on surfaces of atomic tritium weakly bound to  monolayer graphene. Such surfaces are oriented perpendicularly with respect to a high magnetic field, hence electrons are transported through drift processes. Measurements with an  RF antenna (relying on cyclotron radiation emission spectroscopy)  provide an initial estimate of the momentum components of electrons near the endpoint. Then the high-precision and dynamically configured electromagnetic filter takes care of exponentially draining the kinetic energy of electrons. When it reaches a few eV, the residual kinetic energy of electrons is measured by a calorimeter. The combination of measurements with high resolution and the 2D topology aims to guarantee  high precision in the neutrino capture signal. 

The experimental realization was proposed in 2013 with a prototype currently under development at Gran Sasso National Laboratories. PTOLEMY embraces a diverse program from neutrino masses to sterile neutrinos searches to future dark matter detection concepts. It provides an opportunity to open  new insights on neutrino properties, on the physics of the early universe as well as concerning the development of new technologies.

\section{Solar Neutrinos}


Strong support for an ongoing future program in solar neutrinos is critical to ensure that we fully explore the new physics of neutrino mass and oscillations, and that we understand in detail the behavior of our nearest star.

The only significant observed matter effect to date comes from measurements of solar neutrinos.  That observation not only provides the most precise measurement of the mixing angle $\theta_{12}$, which impacts expected rates of neutrinoless double beta decay, but provides excellent sensitivity to new physics in the transition region between matter-enhanced and vacuum oscillations.

Observations of solar neutrinos provide the only significant observed matter effect to-date, which critically impacts our understanding of neutrino properties and behavior, as well as offering a precise handle on mixing parameters that affect the allowed phase space for processes such as neutrinoless double beta decay. This field also offers unique opportunities to probe the solar core, with potential insights into stellar evolution, as well as neutrino interactions and properties (sterile neutrinos, NSIs).

\subsection{Historical context}
Solar neutrinos have a decades-long history of paradigm-shifting results, along with two Nobel prizes, including early demonstrations of neutrino flavor change, and evidence for non-zero neutrino mass.  
Neutrinos are produced by several reactions in the core of the Sun, each giving rise to a characteristic spectrum.  
These neutrinos are associated with two fusion cycles: the \emph{pp} chain, which is the primary source of solar power, and the CNO cycle, which contributes about 1\% of solar power, but offers insights into heavier element production, and the formation of heavier stars.

All neutrinos produced during solar fusion are created in the electron flavour. As they propagate from their production point, they undergo flavour conversion due to the changing matter density of the Sun.  This has provided one of the first opportunities to probe the interaction of neutrinos with matter.


A broad range of experiments have interrogated this source, ranging from a suite of radiochemical experiments that leveraged interactions on either chlorine or gallium -- Homestake, GALLEX, GNO, and SAGE~\cite{1968PhRvL..20.1205D,1998ApJ...496..505C,1992PhLB..285..376A,1999127,2010PhLB..685...47K,2005PhLB..616..174G,PhysRevC.80.015807,2019sone.conf...29G} -- to real-time observations with large-scale optical neutrino detectors leveraging water or liquid scintillator as targets, including Kamiokande, Super-Kamiokande (Super-K), the Sudbury Neutrino Observatory (SNO), KamLAND, and Borexino.
A full chronology of the exciting history of this field can be found in~\cite{Gann:2021ndb}.  Here, we focus on the most recent results, and current open questions.




The Super-Kamiokande (Super-K) experiment made a high statistics measurement of the flux of $^8$B neutrinos via elastic scattering (ES)~\cite{PhysRevLett.86.5651}, which is sensitive primarily to $\nu_e$, but with some admixture of $\nu_x$.  This result was combined with the charged current (CC) measurement from the Sudbury Neutrino Observatory (SNO), an interaction that is sensitive only to $\nu_e$ at relevant energies, allowing a measurement of the pure $\nu_e$ flux~\cite{snocc}.  Disagreement between the two at the 3$\sigma$ level was evidence for some non-electron component in the solar neutrino flux.  This was confirmed by SNO's neutral current (NC) measurement, an interaction that is equally sensitive to all three flavours, and which confirmed at 5$\sigma$ that the $\nu_e$ produced in the Sun's core were changing flavour prior to detection on Earth~\cite{2004PhRvL..92r1301A}.   However, it was the result from the KamLAND liquid scintillator (LS) detector that confirmed that this flavour change was in fact due to oscillation, by observing the characteristic oscillation pattern in the spectrum of reactor neutrinos~\cite{2003PhRvL..90b1802E}.

In more recent years, Borexino has dominated the field of low-energy solar neutrinos, producing a full suite of measurements of neutrinos from almost every branch, including: first direct detection of \emph{pp}~\cite{2014Natur.512..383B} and \emph{pep} neutrinos~\cite{Borexino:pep}, the best precision on the $^7$Be flux~\cite{arpesella:2008,Borexino:7Be}, as well as $^8$B flux and spectrum measurements.  
A comprehensive spectroscopic study of the pp-chain of solar neutrinos followed~\cite{Borexino:phaseI,borexino:pp2,borexino:pp3}.
Perhaps the crowning achievement to date is the first detection of neutrinos from the sub-dominant CNO cycle~\cite{2020Natur.587..577B,Borexino:2022pvu}.

\subsection{Physics opportunities}
A number of open questions remain in the field.  Ongoing experimental programs can offer some insights but, in many cases, new detectors will be needed to continue to push the bounds of our knowledge in this field.

\paragraph{CNO neutrinos}

A precision measurement of CNO neutrinos is one of the most exciting open questions in solar neutrinos.  These neutrinos offer a handle on the metal content of the Sun, which could help to resolve recent discrepancies.  The exact metal content (the metallicity) of a star’s core affects the rate of the CNO cycle. This, in turn, influences the temperature and density profile — and, thus, the evolution — of the star, as well as the opacity of its outer layers.

The metallicity and opacity of the Sun affect
the speed of sound waves propagating through
its volume. 
Historically, predictions from the Standard Solar Model (SSM) agreed beautifully with results from helioseismology regarding the speed of sound in the Sun.
However, more-recent, more sophisticated spectroscopic measurements of photospheric absorption lines
produced results that were significantly lower in opacity than previously thought,
leading to discrepancies with the helioseismological data~\cite{asplund:2005,serenelli:2009}. 
Precise measurements of CNO neutrinos offer the only independent handle by which to investigate this
difference.   Borexino has made the first detection of these neutrinos, but the uncertainty is as yet too large to offer insight on stellar models.  New data is needed, which will require ultra-low backgrounds, low threshold, and excellent resolution in order to improve upon current knowledge.

A recent review~\cite{Gann:2021ndb} considered the prospects for future precision measurements of CNO neutrinos, and found hybrid (Cherenkov + scintillation) detectors (Sec.~\ref{s:hybrid}) to be the optimal approach for percent-level precision, due to the combination of low radioactive background, particle identification, and directional information available in these detectors.



\paragraph{MSW transition}
The current state of understanding of the pp-chain neutrinos can be found in~\cite{Gann:2021ndb}, showing recent results from each experiment; combined fits also exist that further constrain the parameter space~\cite{PhysRevD.94.052010}.  Although data exists in the transition region between vacuum- and matter-dominated oscillation, the precision is not yet sufficient to either confirm the MSW oscillation scenario, or to offer sensitivity to possible NSI models.  Additional data in this sensitive region would offer strong constraints on non-standard models, and further insights into the interaction of neutrinos with matter.


Constraints in the transition region are currently dominated by spectral measurements of $^8$B neutrinos from the water Cherenkov experiments, given the current scale of uncertainty on LS-based measurements in this region.  Super-K offers unparalleled size and access to low-energy neutrinos via the ES measurement, which is highly complementary to the spectral sensitivity from SNO's CC events, albeit with lower statistics, and the flavor-blind NC measurement.
Both Super-K and SNO have contributed to the understanding of the $^8$B spectral shape, pushing the energy threshold to increasingly low energies in order to start to probe the transition region~\cite{leta,2013PhRvC..88b5501A,2016PhRvD..94e2010A}.  Borexino's program has contributed many data points across the full span of the spectrum, and the radiochemical experiments contribute primarily to the vacuum-dominated region.  However, more data is needed in the transition region in order to probe this sensitive regime, to seek to confirm the MSW prediction, and to search for sterile neutrinos, NSIs, and other non-standard effects.


\paragraph{Day/night effect}

Both SNO and Super-K have sought evidence of the predicted day/night effect~\cite{PhysRevLett.112.091805,snodn}, which would lead to some regeneration of $\nu_e$ as neutrinos pass through the earth at night, due to matter effects.   Super-K have measured a nearly 3$\sigma$ indication of this effect.  Borexino have also searched for this effect, and placed strong constraints. 
Greater statistics are required to confirm the precise magnitude and nature of this effect.  

\paragraph{Oscillation parameters}

Reactor measurements of  $\Delta m^2_{21}$ from KamLAND data provide a terrestrial comparison to solar neutrino results, with some small tension persisting in the value of $\Delta m^2_{21}$~\cite{DEGOUVEA2020135751}.  New data is needed to understand whether this discrepancy will persist, and potentially lead to new understanding.  Improved precision on measurements of $\theta_{12}$ will support precision tests of the 3-flavor mixing model, and inform the parameter space for neutrinoless double beta decay searches.

\paragraph{pp neutrinos}
A precise measurement of \emph{pp} neutrinos offers insight into solar luminosity, and the ability to probe the luminosity constraint.
A leading consideration for improved precision in the \emph{pp} measurement is the $^{14}$C background, inherent in any organic scintillator.

\paragraph{hep neutrinos}
A search for \emph{hep} neutrinos is extremely challenging due to the very low predicted flux.  The best limits on this flux come from SNO, and are currently a few times the SSM prediction~\cite{2020PhRvD.102f2006A}.  These neutrinos are the highest energy of the solar neutrino branches, and are produced furthest from the core~\cite{Gann:2021ndb}.  A measurement of this flux could offer insights into the solar density in the outer core.

\subsection{Recent technological developments}

Although a separate topical group (NF10) exists for neutrino detection technology, here we summarise some recent developments that could revolutionize the field of solar neutrinos.

Early water Cherenkov detectors were critical in identifying the solar neutrino signal, thanks to their directional sensitivity, which allows pointing back to the source.  Scintillation experiments in contrast offer high light yields, with associated low thresholds, critical for observation of CNO neutrinos, and excellent resolution.  Particle-depending quenching also offers particle identification (PID) via pulse shape discrimination. 
An experiment that can combine these features would have unprecedented sensitivity to solar neutrinos, allowing enhanced background rejection with good resolution, at low threshold.

Although existing LS experiments were not designed with this capability in mind, significant progress has been made towards realizing directional sensitivity in these detectors.  The Borexino Collaboration continues to break ground with their recent result that demonstrated statistical direction reconstruction using early-time PMT hits~\cite{BOREXINO:2021efb}.  SNO+ has indications of event-level directional sensitivity in LS data, enhanced by the current low fraction of PPO, which results in a slower scintillation time profile, enhancing the clarity of the prompt Cherenkov component.

Significant work is ongoing within the community to develop an experiment that would be a truly ``hybrid'' detector -- designed to leverage both Cherenkov and scintillation light together.  The potential of this technology for solar neutrinos is discussed in Section~\ref{s:hybrid}.  The relevant technology, which includes novel scintillators, fast photon detectors, and spectral sorting, is discussed in the report from NF10.

Another novel technology that is under development is opaque liquid scintillators.  These intentionally shorten the scattering length of the LS in order to confine light near the interaction point, thus providing excellent vertex resolution and information regarding event topology.  These are discussed in Section~\ref{s:LO} and, again, the technology development is described in more detail in the report from NF10.

New developments in photon detection techniques could also prove critical to this field.  Fast photon detectors, such as Large Area Picosecond Photon Detectors (LAPPDs) and spectral sorting techniques, such as dichroicons, offer the potential for significantly enhanced precision for low-energy neutrino detection.  A solution to large area, low noise, high quantum efficiency, cost effective optical detectors would be game changing across the field.

\subsection{Prospects for future measurements}
\subsubsection{Water Cherenkov detectors}
\subsubsubsection{Super Kamiokande}

Super-K has benefited from a number of upgrades over its lifetime.  The most recent of these is the addition of gadolinium, a project known as SK-Gd, which will enhance neutron capture efficiency~\cite{PhysRevLett.93.171101,Takeuchi:2020slv}.  This can improve the separation of solar neutrinos from radioactive background from cosmic ray induced spallation, improving Super-K's sensitivity to the day/night effect and the $^8$B spectral shape.

\subsubsubsection{Hyper Kamiokande} 

The sheer size of the Hyper-Kamiokande (Hyper-K) experiment will make it a powerful tool for solar neutrino measurement. 
The ability to reconstruct the charged particle's direction also offers a robust signal/background discriminant.  
Construction and operation is expected to begin this decade~\cite{HK}.  At over 250 ktonne total mass, with 40\% coverage, 
Hyper-K has the potential to contribute a great deal to the picture of high-energy solar neutrinos, in particular.  

Hyper-K will detect ES of $^8$B neutrinos, offering sensitivity to the spectral shape and the day/night effect.  
The former in particular is a statistically-limited measurement, making Hyper-K the perfect tool to provide improved precision.
Within 10 years of operation, Hyper-K will be able to measure the day/night asymmetry to better than 4 (8) $\sigma$ at the values currently predicted by reactor (solar) experiments. 

Sensitivity to non-standard interactions via a spectral measurement of $^8$B in the vacuum-matter transition region will depend on the energy threshold achieved.  At 3.5 (4.5) MeV, 5 (3) $\sigma$ sensitivity is predicted to the so-called ``MSW upturn'' predicted by the standard picture of  oscillations.  Deviation from that predicted shape could suggest the presence of NSIs or sterile neutrinos.

The high statistics available to Hyper-K will make it possible to observe short time-period variations in the solar neutrino flux, resulting in something akin to real-time monitoring of the solar core, and may also facilitate the first significant detection of \emph{hep} neutrinos.

\subsubsection{Liquid scintillator detectors}

\subsubsubsection{SNO+}

SNO+ is a currently operating, kton-scale LS detector.  Located in SNOLAB in Ontario, Canada, SNO+ benefits from one of the deepest sites available for low-background studies, at 6 km water equivalent.  This results in incredibly low cosmogenic backgrounds, in particular the $^{11}$C that can be a limiting factor in precision low-energy solar neutrino measurements.
The primary goal for SNO+ is a search for neutrinoless double beta decay (NLDBD) via loading of the LS with tellurium, but during the current pure LS operations, SNO+ will have sensitivity to solar neutrinos, in particular the $^8$B neutrinos~\cite{Andringa:2015tza}.   
Preliminary LS data from SNO+ shows levels of radon daughters that, although sufficient for the NLDBD target, may limit the low-energy solar neutrino program. 
Precision measurements of this regime would require significant reduction of these backgrounds, similar to that achieved during the first period of Borexino operations~\cite{arpesella:2008}. This could potentially be addressed by methods that would include recirculating through the SNO+ scintillator processing systems, built with several purification capabilities, along with other background reduction techniques. 

Early data from the initial water phase of SNO+ have already demonstrated low levels of cosmogenic and external background, allowing a measurement of the $^8$B spectrum in water~\cite{2019PhRvD..99a2012A}.  Sensitivity to CNO neutrinos can be estimated under certain assumptions for background reduction.  With background reduction of approximately a factor of 10 for the U- and Th-chains and 1000 for $^{210}$Bi, negligible $^{40}$K, and constraining the \emph{pep} flux based on the \emph{pp} flux, as was done by Borexino in their discovery paper, SNO+ could achieve better than 15\% precision on the CNO flux, which could be sufficient to resolve the solar metalicity question. 

SNO+ will also offer improved precision in $\Delta m^2_{21}$.  Sensitivity to this parameter using both solar neutrinos and reactor neutrinos will provide additional data to resolve current (small) discrepancies in measurements from solar and terrestrial sources.  

\subsubsubsection{JUNO}

The Jiangmen Underground Neutrino Observatory (JUNO) is under construction in China, with a 20-kton target mass and a goal of 3\% energy resolution at 1~MeV~\cite{JUNO,JUNOCDR}.  Although the primary goal is a precision reactor neutrino measurement, the 700-m overburden and planned radiopurity mean JUNO will have a rich physics program, including certain solar neutrino measurements.  The relatively shallow site will limit the low-energy solar neutrino program, but the large size and precision resolution will offer excellent sensitivity to both $^7$Be and $^8$B neutrinos, allowing some insights into solar metalicity.
At the target background levels, a threshold of 2~MeV could be achieved for $^8$B neutrinos.  This is substantially lower than what is possible in a Cherenkov detector, and 1 MeV lower than achieved by Borexino for a measurement of the $^8$B spectral shape~\cite{PhysRevD.82.033006}.  This spectral measurement gives JUNO sensitivity to non-standard interactions, and 2 (3) $\sigma$ sensitivity to the day/night effect at current reactor- (solar-) favoured parameter values~\cite{JUNO8B}.  With the ability to measure $\Delta m^2_{21}$ to percent-level precision from reactor neutrinos, and to approximately 20\% using solar neutrinos, JUNO will provide a uniquely precise cross check on the consistency of data from these two sources.  

\subsubsubsection{LiquidO}\label{s:LO}
The use of opaque scintillator allows confinement of the light from a neutrino interaction close to the original interaction point~\cite{Buck:2019tsa}.  Combined with fibre instrumentation, this results in an event imaging capability that can enhance background rejection even at low energies.  The clean signatures of different particle types allow for separation of electrons from gammas, offering excellent rejection of radioactive background for solar neutrino detection.  The possibility of doping also opens up the option of charged current detection, for example on indium, allowing for excellent spectral sensitivity~\cite{Cabrera:2019kxi}.

\subsubsection{Hybrid optical detectors}\label{s:hybrid}
So-called ``hybrid'' optical neutrino detectors seek to leverage the combined benefits of Cherenkov and scintillation light in a single detector.

\subsubsubsection{Slow fluors}
By deploying certain fluors in combination with LS, the time profile of the scintillation can be delayed~\cite{MinfangSlow,BillerSlow}, thus enhancing the ability to separately identify Cherenkov photons, and to reconstruct direction for low-energy interactions.  
This improvement must be balanced against potential degradation in vertex reconstruction due to reduced precision in photon time-of-flight information.  However, optimisation can lead to 10\% sensitivity to the CNO neutrino flux from a few kton-yr exposure in a high light yield LS detector.

\subsubsubsection{\theia }

\theia is a proposed large hybrid optical neutrino detector, capable of leveraging both Cherenkov and
scintillation light in a single detector.  \theia is a realisation of the Advanced Scintillation Detector Concept first proposed in~\cite{asdc}.
This concept offers a number of benefits, including
directional sensitivity at low threshold, combined with the good vertex and energy resolution of a scintillator detector, as well  as 
further handles on particle and event identification -- in addition to the PSD-based PID traditionally used in LS detectors -- from the Cherenkov / scintillation ratio. 
\theia would combine a novel LS target - such as water-based LS (WbLS) or a slow scintillator, with fast, high efficiency photon detectors and potential spectral sorting using technology such as dichroicons~\cite{Kaptanoglu:2019gtg} to optimize this hybrid detection capability. The proposed site at the Sanford Underground Research Facility (SURF)  laboratory in SD, USA would offer a high-energy neutrino programme as part of the Long Baseline Neutrino Facility, as well as a broad program of low-energy physics.  Although pure LS offers improved radiopurity, directional sensitivity would at least in part offset the increased levels of contamination inherent in the water component of a WbLS target.
\theia offers better than 10 (1)\% precision on the CNO solar neutrino flux with
a WbLS (pure LS) target~\cite{Bonv,theiawp,MeVSolar}, allowing a high-confidence resolution of the metallicity problem, as well as sensitivity to the shape of the low-energy 8B spectrum, providing sensitivity to non-standard effects in the vacuum-matter transition region.
Possible isotope loading is also being explored to offer a CC interaction, which would provide a high fidelity measure of the underlying neutrino spectrum, potentially offering enhanced sensitivity both to CNO neutrinos, and to the $^8$B spectral shape. 

The power of hybrid detection means that even a relatively small, few-hundred ton fiducial detector can achieve enticing reach if filled with a high light yield, ultra clean scintillator, as explored in~\cite{MeVSolar}. With current technology: linear alkyl benzene (LAB) with 2~g/L of PPO, and instrumented with standard photomultiplier tubes, at 1.6-ns transit time spread, such a detector could achieve 14\% precision on the CNO neutrino flux, dropping below 10\% if a constraint is imposed on the pep flux based on knowledge of the pp flux, as done by Borexino in their discovery paper~\cite{2020Natur.587..577B}.  With fast photon detectors, such as LAPPDs~\cite{LAPPD}, the precision is below 5\%.

\subsubsubsection{Other underground LS detectors}
A few-kton hybrid LS detector is under consideration for deployment at the China Jinping Underground Laboratory (CJPL).  The 2.4-km rock overburden results in a cosmic-ray muon flux almost as low as that at SNOLAB~\cite{JinpingMuons}.   
Detector sensitivity depends on the final configuration, with options from 1- to 4-ktonne in fiducial mass, and 200 to 1000 photoelectron/MeV light collection under consideration~\cite{JinpingLOI}.  The optimal detector would achieve percent-level measurements of \emph{pp}, $^7$Be, and $^8$B neutrinos, and a few-percent on the \emph{pep} flux.  A CNO measurement could reach better than 15\% precision for the larger, high-resolution detector configurations.  The experiment would have good sensitivity to the $^8$B spectral shape. 
A 100-ton prototype is planned for the middle of this decade.  


A 20-m right cylindrical detector has been proposed in YemiLab, in Korea, and a funding proposal is currently under consideration, with a White Paper under preparation.  Such a detector would consider of over 2~kton of LS, with large-area PMTs for photon detection.  This detector would interrogate a number of low-energy sources, including solar neutrinos.

A new underground laboratory is being developed in the Agua Negra tunnel: ANDES (Agua Negra Deep Experiment Site) between Argentina and Chile.  This would be the first deep underground laboratory in the southern hemisphere, and could provide significant opportunities for deep underground experiments, such as precision solar neutrino searches.

\subsubsection{Noble liquid and solid state detectors}
\subsubsubsection{Nuclear recoil detectors}
Experiments designed to search for coherent neutrino-nucleus elastic scattering, known as CE$\nu$NS~\cite{cevns,cevns2}, can offer sensitivity to higher-energy solar neutrinos via this channel.  The cross section for CE$\nu$NS interactions is favourable due to an A$^2$ dependence; however, the nuclear scatters typically fall below 10~keV, requiring detectors with excellent resolution and very low threshold. CE$\nu$NS offers precise flux measurements for the higher energy fluxes, and potential spectral sensitivity across the range from vacuum- to matter-dominated oscillation, giving sensitivity to possible active-to-sterile mixing~\cite{sterile1409}.

Technologies include scintillating crystals, germanium semiconductors, and noble liquid detectors.  Some of these offer parallel sensitivity to an ES signal as well.  The COHERENT suite of detectors will include several of these, and will make measurements of CC and NC cross sections of relevance for future CE$\nu$NS detection of solar neutrinos.  A more detailed understanding of these interactions will be important for interpretation of data from these future experiments.

Ultra low-threshold detectors such as SuperCDMS could potentially detect the low-energy branches, such as \emph{pp} neutrinos, if thresholds in the few-eV range can be achieved~\cite{supercdms,nucleus}.  However, equaling the precision of existing measurements from Borexino would require ton- to multi-ton-scale exposure~\cite{LouisLowth}.

CYGNUS is a proposed gaseous He / SF$_6$ TPC, which would have directional sensitivity to nuclear recoils.  A 1000 m$^3$-scale experiment could detect 10s of solar neutrinos, with a large detector being needed for precision measurements.

Noble liquid bubble chambers offer low threshold sensitivity to nuclear recoils and, to-date, relative insensitivity to electron recoils, providing good background discrimination even at low, sub-keV energies.  This technology could be used for CE$\nu$NS detection of solar neutrinos down to a threshold of 1.4~MeV.  A 10-kg experiment will be deployed at SNOLAB as a demonstration.  A future, ton-scale or larger detector could have sensitivity to detect up to 40 NC events from $^8$B interactions per ton-yr.


\subsubsubsection{Liquid xenon}
A Generation-3 liquid xenon (LXe) dark matter (DM) detector, such as DARWIN or a future PandaX upgrade, will have sensitivity to solar neutrinos via both nuclear and electron scattering.  Such a detector combines an extremely low threshold and good radiopurity, with excellent PSD for discrimination of electron and nuclear interactions.  This offers a broad program of solar neutrino physics: precision \emph{pp} flux measurement via ES; 1.4\% precision on a measurement of $\sin^2\theta_W$; flux and spectral measurement of $^8$B neutrinos, for sensitivity to NSI and sterile neutrinos; a measurement of the CNO neutrino flux, if the $\nu\nu\beta\beta$ background can be reduced by 3 orders of magnitude; and sensitivity to \emph{hep} neutrinos via CE$\nu$NS, if resolution can be improved in the 1--3~keV$_{nr}$ range.

The PandaX observatory currently consists of a 4-ton LXe TPC, located in the CJPL.  With anticipated low background in the range from keV to 10~MeV, a future upgrade with a 30--100 ton target mass could perform a precision measurement of the \emph{pp} solar neutrino spectrum, allowing a sensitive search for an anomalous neutrino magnetic moment.  This detector could also observe coherent nuclear scattering of $^8$B neutrinos, offering sensitivity to NSIs.

\subsubsubsection{DUNE, and other liquid argon TPCs}
The Deep Underground Neutrino Experiment (DUNE) will deploy a number of liquid argon (LAr) TPC modules at the Sanford Underground Research Facility, to view the neutrino beam from Fermilab. While the primary goal is to search for CP violation, this detector may also offer sensitivity to higher-energy solar neutrinos.  Although it may be limited by intrinsic radioactivity in the detector, DUNE could offer improved measurements of $\Delta m^2_{21}$ as well as measurements of the $^8$B and \emph{hep} fluxes.

A DUNE-like module with a focus on achieving low background, via underground argon, enhanced radiopurity requirements, additional shielding, and improved coverage and readout for enhanced in-situ background discrimination --  would expand the solar neutrino potential to include \emph{hep} neutrinos, and lower-energy fluxes such as \emph{pp} neutrinos~\cite{Caratelli:2022llt}.  Sensitivity to solar oscillation parameters and the shape of the survival probability, via spectral measurement, would allow searches for NSIs, as well as to address the current mild tension between solar and terrestrial measurements of oscillation parameters.  Some studies consider the use of the Cherenkov signal in LAr as a handle for background discrimination, which could enhance detection of the solar ES signal.  Sufficient reduction of $^{42}$Ar might even make a measurement of CNO feasible in such a detector~\cite{https://doi.org/10.48550/arxiv.2203.08821}.

The Global Argon Dark Matter Collaboration (GADMC) plans to deploy purified LAr in a range of TPCs for DM detection.  Solar neutrino sensitivity is offered via both CE$\nu$NS and CC interaction on argon.  A 1500 ton-yr exposure would allow collection of several thousand CNO neutrino interactions, along with 6000 \emph{pep} events and 16000 from $^7$Be.  Flux measurements will require improvements in radon reduction by 3 orders of magnitude beyond that achieved in smaller experiments~\cite{LArTPC}.  CE$\nu$NS measurements of $^8$B and, if use of underground argon allows the threshold to be pushed below 200~eV$_{nr}$, \emph{pep} and CNO neutrinos will also be accessible.

\subsubsubsection{Selena}

Designed as a search for neutrinoless double beta decay, the Selena experiment could also offer sensitivity to solar neutrinos via electron neutrino capture on $^{82}$Se~\cite{Chavarria:2021rbw}.  Selena will detect ionization on a pixelated array read out via CMOS.  Track reconstruction allows for particle identification, and spatio-temporal correlation allows for identification of coincidence decays even over extremely long half lives. This strategy targets background-free measurements of solar neutrinos.  The 172~keV threshold for electron neutrino capture offers sensitivity to all branches of solar neutrinos.  The resulting signal offers a three-fold coincidence that results in negligible accidental background and, to-date, no identified background isotope that can mimic the particular decay sequence.  A 100 ton-year exposure would provide 1\% precision on \emph{pp} neutrinos, 8\% for \emph{pep}, and 10\% on the CNO neutrino flux.  Measurement of the energy of $^7$Be neutrinos, and of the predicted shift from the energy of the equivalent laboratory decay, offers sensitivity to the solar core temperature at the 25\% level.

\subsubsection{Other concepts}
\subsubsubsection{High energy solar neutrinos}
High-energy neutrinos can be produced from dark matter (DM) annihilation, and from interaction of cosmic rays with the solar atmosphere.  To-date, both searches have yielded null results, but include highly-competitive limits on the spin-dependent DM-nucleon cross sections for DM with mass above $\approx$100 GeV.  These searches are statistically limited, and larger next-generation detectors, including Baikal-GVD, Hyper-Kamiokande, IceCube Gen2, and KM3Net can add to the picture.  First detection of 
solar atmospheric neutrinos is within the reach of next-generation detectors, and would allow neutrino telescopes to explore new baselines, and matter densities.
High-energy neutrinos can also be produced in solar flares, but have yet to be detected.

The idea of deploying a neutrino detector on a spacecraft offers scope for a smaller detector at shorter baselines, and the ability to vary the point of observation.  Such a project would be costly, and would need to travel extremely close to the Sun to achieve sufficient detection rates.



\section{Geoneutrinos}

The reader is referred to~\cite{Bellini:2021sow} and~\cite{Sramek:2012hma} for a comprehensive description of the geological questions and associated experimental program in this area.  The following section attempts to provide an overview of the current status of the field, with a focus on prospects in the next decade.

\subsection{Motivation}
Radioactive decay of naturally occurring uranium, thorium and potassium in the Earth accounts for over 99\% of radiogenic heat production.  These decays produce a flux of antineutrinos, so-called ``geoneutrinos'', that is proportional to the heating power. Measurements of geoneutrinos offer us the ability to assay the earth, to probe radiogenic heat production, and test models of the Earth's composition.

Questions that can be addressed using geoneutrinos include:
\begin{itemize}
    \item Understanding the contribution of radiogenic (U, Th) heat production to heat flow and energetics in the deep Earth.
    \item Understanding the degree to which mantle convection is driven by radiogenic heat.
    \item Testing models of the crust composition.
    \item Potentially, probing the distribution of reservoirs in the mantle, the nature of the core-mantle boundary, and the existence of radiogenic elements in the core.
\end{itemize}

Measurements of geoneutrinos can shed light on these questions by measuring the total geoneutrino flux, and the U/Th ratio, inferred from a spectral fit.  Geographical variations in the flux and U/Th ratio are also of great interest, since these can be used to constrain the relative contributions from the crust and mantle independently from model constraints.

\subsection{Current understanding: Borexino and KamLAND}

Two experiments to date have successfully measured the geoneutrino flux: the KamLAND experiment in Japan, and Borexino in Italy.  Both experiments detected the inverse beta decay (IBD) of a electron antineutrino on a proton, yielding a positron and neutron.  The coincidence signature allows for suppression of many sources of background, and the high light yield of the chocen liquid scintillator (LS) technology allows for a spectral fit, which can be used to extract limits on the U/Th ratio.  

Results from these experiments have demonstrated the technique, and yielded results consistent with model predictions, albeit with large uncertainties.  The precision is not yet sufficient to determine the U/Th ratio.  Additional, and more precise measurements will be important to facilitate extraction of geological results in the future.

KamLAND was built to detect antineutrinos produced from nuclear reactors, and has a long history of successful antineutrino physics measurements.  Backgrounds to a geoneutrino search are dominated by reactor antineutrinos, due in part to the overlap in the energy spectra of the two signals.  The comprehensive unbinned maximum likelihood analysis also takes into account cosmogenic backgrounds, fast neutrons, atmospheric neutrinos, ($\alpha$,n) reactions, and accidental coincidences~\cite{Araki:2005qa,PhysRevD.88.033001}.  The oscillated electron antineutrino flux from the sum of $^{238}$U and $^{232}$Th measured at the Earth's surface was found to be $3.6^{+0.6}_{-0.6}\times 10^6$cm$^2$s$^{-1}$.  The number of geoneutrinos observed was $168.8^{+26.3}_{-26.5}$.  This 15.6\% uncertainty is a result of fixing the Th/U mass ratio in the fit.  The absence of geoneutrinos is rejected at over 8$\sigma$.  Full details of the analysis, and results for the case of an unconstrained Th/U ratio can be found in~\cite{Bellini:2021sow}.

The Borexino experiment has executed an extensive program of low-energy solar neutrino measurements, thanks in part to its excellent levels of radio purity, and well understood detector response.  These features also support an antineutrino program, and Borexino has successfully measured the geoneutrino flux at LNGS in Italy. As for KamLAND, the dominant background is from reactor antineutrinos, the flux of which is fit for in the analysis and found to be consistent with expectation.  Over 3262.74 days of data, a signal of $52.6^{+9.6}_{-9.0}$ geoneutrinos was observed from the sum of $^{238}$U and $^{232}$Th.  The resulting precision is approximately 18\%, with a constrained Th/U mass ratio.

These results can also be interpreted in the context of various geological models.  By constraining the crust contribution, the null hypothesis of geoneutrinos from the mantle can be excluded at 99\% C.L.  The resulting signal can be converted into a radiogenic heat of $24.6^{+11.1}_{-10.4}$ TW from $^{238}$U and $^{232}$Th in the mantle.  Combining this with certain values from model predictions, the total radiogenic heat of the Earth can be estimated, and compared to a range of model predictions.  Although consistent with all models to within 3$\sigma$, a preference is
found for models with relatively high radiogenic power, corresponding to a cool initial environment at Earth’s early formation stages, and small values of
heat from secular cooling.
By fitting the data with a constraint on the reactor antineutrino background, the existence of a hypothetical georeactor at the center of the Earth having power greater than 2.4 TW at 95\% C.L. can be excluded. A comprehensive description of this analysis can be found in~\cite{PhysRevD.101.012009}, and a summary in~\cite{Kumaran:2020pfs}.

\subsection{Prospects for future measurements}
Future measurements of geoneutrinos will seek to increase statistics, add data points at new geographical locations, and improve precision in order to enhance sensitivity to the Th/U ratio via a spectral analysis.  \cite{Bellini:2021sow} provides a comprehensive overview of several forthcoming projects.  The following summarises what each project could achieve for geoneutrino measurement.

It is important to emphasise that significant power comes from combined analyses of data from experiments at different geographical locations~\cite{Fogli:2010vx}.  Each experiment needs the local crust geology to be characterised, such that a combined analysis yields the common mantle component, or becomes sensitive to potential lateral inhomogeneity in the mantle component.  It will be important to support effort towards such analyses in order to maximally benefit from forthcoming experiments.

\subsubsection{SNO+}
SNO+ is an operational LS experiment in Sudbury, Canada.  At 780 tons of pure LS, SNO+ can expect to observe $25--30$ geoneutrino events per year, with a relatively low reactor background: approximately 1:1 in the region of interest.  The local geology has been extensively characterised, and the region is predicted to have a higher natural flux than at either KamLAND or Borexino, making SNO+ a useful component in a future global analysis.

\subsubsection{JUNO}
The JUNO experiment is under construction in China, with a planned 20-kton LS volume and excellent energy resolution, of 3\% at 1~MeV, in order to target a hierarchy measurement from a precision measurement of the oscillated reactor neutrino spectrum.  The predicted geoneutrino signal is approximately 400 events / yr, with a reactor background of approximately 8 times that in the relevant region of interest.  This should allow for a 5\% precision measurement of the geoneutrino flux with 10 years of data, giving approximately 30\% uncertainties on the Th/U ratio~\cite{Han:2015roa}.

\subsubsection{Ocean Bottom Detector (OBD)}
The crustal component of the geoneutrino flux dominates all measurements to date.  The oceanic crust is thin, with low Th and U abundances, allowing greater sensitivity to the mantle contribution.  A detector in the ocean could meet many of the needs of a geoneutrino program.  Such a detector could be moved to multiple locations, and would observe a signal dominated by the mantle contribution.  With appropriate detector technology, this detector could measure the total radiogenic heat in the Earth, the Th/U ratio, distinguish the mantle contribution, and potentially resolve flux differences at different locations.  Originally conceived of as the Hanohano experiment, an OBD could be a powerful instrument for physics, offering a broad program of low-background astro-particle physics complementary to the geoneutrino measurement~\cite{OBD}.  
A prototype liquid scintillator detector has been funded, as part of a broader program to develop and deploy a multi-kton detector, and efforts are ongoing to deploy this detector on the sea floor.
Monte Carlo studies show that a 1.5-kton scale detector could achieve better than 3$\sigma$ on a mantle geoneutrino measurement with 3 years of data~\cite{Sramek:2012nk}.

\subsubsection{\theia}
\theia is a proposed large-scale (10s of kton) ``hybrid'' optical neutrino detector, capable of leveraging both Cherenkov and scintillation light simultaneously~\cite{theiawp}.  The addition of the scintillation component improves energy and vertex reconstruction and, critically for a geoneutrino measurement, allows a lower threshold than a traditional water Cherenkov detector.  However, the ability to detect the Cherenkov signal offers some handle on direction reconstruction, and additional particle identification -- beyond the PSD-based PID traditionally used in LS detectors -- from the Cherenkov / scintillation ratio.  These features can improve background rejection.
\theia would offer a very high statistics measurement in the continental US, with the potential to extract the Th/U ratio from a spectral fit.

\subsubsection{Other underground LS detectors}
The Jinping underground laboratory in China offers an extremely deep environment for low-background and rare event searches.  A few-kton liquid scintillator experiment is being designed for this space, which could offer good sensitivity to the geoneutrino flux. Although to-date there is no US involvement in this effort, it can contribute to the global picture of geoneutrino measurements, and US efforts can benefit from this complementarity.

A future detector at YemiLab in Korea, or in the proposed ANDES site in South America, could also offer the potential for measurements of geoneutrinos.

\subsubsection{The search for $^{40}$K geoneutrinos}
Below the threshold for IBD detection, $^{40}$K geoneutrinos are expected to contribute significantly to the overall flux, but offer a unique challenge for detection.  One approach involves ES detection in a Cherenkov-sensitive LS detector (a so-called hybrid detector), such as those discussed above -- \theia, JUNO, Jinping, or another kton-scale or larger detector~\cite{Wang:2017etb}.

\subsubsection{LiquidO}
The LiquidO concept involves a highly instrumented volume of opaque scintillator, designed to confine light close to the interaction point, thus preserving information about the vertex.  This imaging capability offers a degree of PID that enhances background rejection even at low energies.  The possibility of doping also opens up new windows, such as a charged current detection of geoneutrinos, with a lower threshold than IBD, potentially allowing detection of the $^{40}$K signal.

\subsubsection{Coherent neutrino-nucleus scattering}
Coherent elastic neutrino-nucleus scattering allows sensitivity to low-energy neutrinos.  A diverse array of detector technology can be employed for this purpose, from scintillating crystals, to germanium semiconductors, to noble liquid detectors.  This technique may offer sensitivity to the geoneutrino flux, although such detectors would need to combat the large background from solar and reactor neutrinos, and the energy of recoils from the geoneutrino signal would be extremely low~\cite{PhysRevD.99.093009}.  Nuclear recoil detectors with directional sensitivity, such as CYGNUS, could offer an additional handle for this~\cite{geodirarticle}.

\subsubsection{$^6$Li-doped scintillator} 
$^6$Li doping of organic scintillators offers many benefits for antineutrino detection, including excellent PSD for n/g discrimination, large neutron capture cross section, and high light yield.  Improved precision in the IBD detection from such techniques could enhance directional sensitivity, which would potentially allow for separation of different components of the antineutrino signal.

\section{Supernova  Burst Neutrinos and Diffuse Supernova Neutrino Background}

\subsection{Motivation and current understanding}

Core-collapse supernovae stem from collapsing massive stars, during the collapse about $10^{58}$ neutrinos are emitted for $\mathcal{O}(10)$~s with average energy of $\mathcal{O}(10)$~MeV~\cite{Vitagliano:2019yzm, Mirizzi:2015eza, Janka:2017vcp, Burrows:2020qrp}. Being produced so abundantly, neutrinos play a crucial role in  supernovae, transporting energy and lepton number and giving insight into the extreme environment within the interior of the progenitor star. 

The majority of supernovae are assumed to explode because neutrinos revive the stalled shock wave to finally trigger the explosion. While major progress has been made in the context of hydrodynamical simulations of the core collapse~\cite{Mezzacappa:2020oyq,Janka:2017vcp, Burrows:2020qrp}, flavor mixing is neglected in the modeling of neutrino transport in hydrodynamical simulations.  In addition,  magneto hydrodynamical simulations are yet to include sophisticated neutrino transport. 

The supernova neutrino signal contains an imprint of both the supernova environment, with extreme conditions that we cannot recreate in the  laboratory, and the properties of the neutrinos themselves. Neutrinos free-stream from the supernova, acting as an early warning for electromagnetic telescopes with advanced notice between minutes and days \cite{Kistler:2012as}. This early warning will enable observations of the first light from the shock breakout, carrying important progenitor and explosion properties, such as the radius and surface composition, as well as the mass-loss history \cite{Waxman:2016qyw}. 

The detection horizon for the burst of neutrinos from a core-collapse supernova with current detectors is limited to our own galaxy and the surrounding satellite galaxies. Hyper-Kamiokande will extend the horizon out to Andromeda, our neighboring galaxy. Because we are limited to nearby stars, the rate of core-collapse supernovae from which we can expect to detect neutrinos is around a few per century \cite{Rozwadowska:2020nab}. It is therefore imperative to have a continuous observing campaign using neutrino detectors with high uptime in order to capture this once-in-a-lifetime event. 

\subsubsection{Neutrino flavor mixing}


In order to extract the physics from the supernova signal, theoretical progress is needed. Despite huge advancements occurring in the past decade concerning  the development of multi-dimensional hydrodynamical simulations with sophisticated transport of neutrinos~\cite{Mirizzi:2015eza,Burrows:2020qrp,Mezzacappa:2020oyq,Janka:2017vcp}, our understanding of  neutrino flavor conversions in supernovae remains preliminary~\cite{Mirizzi:2015eza, Duan:2010bg, Tamborra:2020cul,Chakraborty:2016yeg}. In fact, because of the large density of particles, the coherent forward scattering of neutrinos onto each other is responsible for making the flavor evolution non-linear; moreover,   the latter is crucially affected by the neutrino angular divergence. A relatively recent development concerns the occurrence of flavor mixing triggered by the pairwise scattering of neutrinos in the proximity of the weak decoupling region in the supernova core~\cite{Tamborra:2020cul,Chakraborty:2016yeg,Richers:2022zug}. Because of the numerical complications intrinsic to the modeling of this non-linear phenomenon, a full modeling of flavor mixing is lacking; hence, the consequences of neutrino mixing on the observable neutrino signal remain to be clarified as well as its effect on the explosion dynamics itself.

\subsubsection{Neutrinos as probes of the supernova physics}

Despite the uncertainties related to the  neutrino physics, the neutrino signal could be a crucial  probe of the supernova physics taking place just before the explosion~\cite{Tamborra:2013laa,Walk:2019miz, Walk:2018gaw,Muller:2019upo}. Neutrinos are also emitted in the latest stages preceding the stellar collapse~\cite{Patton:2015sqt,Patton:2017neq}; their detection for nearby core collapses will be crucial to inform astronomers about the upcoming supernova burst and to provide insight on the pre-collapse phase.
In addition, the neutrino signal can be adopted to optimize the time window  for the search of gravitational waves~\cite{Nakamura:2016kkl,Adams:2013ana} and will be a crucial observable in the case of black hole forming collapses~\cite{OConnor:2010moj}.  The long term neutrino emission during the cooling phase may help to extract information on  the nuclear equation of state and neutron star formation as well as the nucleosynthesis~\cite{Li:2020ujl, GalloRosso:2018ugl,Arcones:2012wj, Cowan:2019pkx}. 

\subsubsection{Supernova pointing and early warning through neutrinos}

Neutrinos from a supernova will arrive on Earth minutes to days before the photons, and therefore will provide a valuable early warning for multi-messenger followup. The direction of the supernova can be extracted using the electron elastic scattering detection channel, which can preserve the direction of the incoming neutrino. Super-Kamiokande has the best prospect for this measurement due to its large volume and because detection via the Cherenkov light preserves direction. The expected resolution in Super-Kamiokande is between $\sim$ 3 and 6 degrees for a supernova at 10 kpc. As more detectors come online in the near future, using triangulation to compare the arrival times of the signal in detectors around the globe can yield directional constraints. In \cite{Linzer:2019swe,Sarfati:2021vym}, the arrival of the early signal to triangulate the direction was found to constrain the direction to a few percent of the sky in the most favorable case. Fitting the shape of the neutrino curve can further improve the uncertainty of arrival time in detectors \cite{Brdar:2018zds}. 

Measuring neutrinos preceding collapse can give advance warning to neutrino and gravitational wave detectors that the burst of supernova neutrinos is imminent. These neutrinos are lower in energy and therefore liquid scintillator experiments are ideally suited for this measurement. Sensitivity to these presupernova neutrinos was first demonstrated by Kamland \cite{KamLAND:2015dbn}. The recent addition of Gadolinium to Super-Kamiokande has also allowed for a reduced energy threshold and a corresponding sensitivity to the presupernova signal \cite{Super-Kamiokande:2019xnm}.  Future large-scale dark matter experiments are also expected to have sensitivity \cite{Raj:2019wpy}.

\subsubsection{Neutrinos as probes of physics beyond the Standard Model}
In the light of current uncertainties on the physics of neutrino flavor conversion and related degeneracies with  the supernova properties (e.g., its mass, nuclear equation of state, etc.), the neutrino signal from the next nearby supernova burst is unlikely to provide smoking gun signatures on the mass ordering of neutrinos and mixing parameters. However, New Physics may largely modify the detectable neutrino signal as well as the supernova physics itself. Of special interest in this context are New Physics scenarios directly linked to neutrinos or leading to lepton number violation~\cite{Suliga:2019bsq,deGouvea:2019goq, Das:2017iuj,Nunokawa:1997ct,Tamborra:2011is,Shalgar:2019rqe, Carenza:2020cis, Stapleford:2016jgz,Berryman:2022hds}.

\subsubsection{Current experimental landscape}
Current neutrino detectors will be able to detect a burst of neutrinos from a core-collapse supernova if sufficiently nearby, in our galaxy or the surrounding Magellanic clouds. The time-dependent rate of supernova neutrinos will allow us to produce a neutrino lightcurve, on which will be imprinted the details of the explosion and the dense environment. The statistics for the lightcurve will be driven by water and in-ice Cherenkov detectors, which primarily detect supernova neutrinos via inverse beta decay. The large instrumented volume of IceCube results in the best statistics to measure the neutrino lightcurve for a supernova in our galaxy \cite{IceCube:2011cwc}. Super-Kamiokande, will measure supernova neutrinos on an event-by-event bases allowing not only a study of the neutrino lightcurve but also the time-evolution of the neutrino energy \cite{Super-Kamiokande:2007zsl}. 

Flavor-independent measurements of supernova neutrinos are of great importance to disentangle the complicated flavor-mixing effects arising in supernovae. Liquid scintillator detectors such as SNO+ or JUNO may have sensitivity to the neutral current interaction on Carbon. Detectors using heavy nuclei such as lead, including the currently-operating HALO experiment, would expect a modest number of interactions via neutral current interactions. A particularly interesting development in the detection context  concerns the employment of coherent neutrino nucleus scattering~\cite{Drukier:1984vhf}. In recent years, there has been growing attention on the possibility of detecting supernova neutrinos through direct detection dark matter experiments~\cite{Lang:2016zhv} as well as proposals to build  neutrino telescopes exploiting this detection channel~\cite{Pattavina:2020cqc}. 

\subsection{Future Prospects}

\subsubsection{Prospects for theoretical developments}

As existing and upcoming neutrino observatories prepare for the next supernova burst, the theory community should progress in the modeling to be ready to take full advantage of the precious information carried by the supernova explosion. The fact that neutrinos through their flavor mixing could radically affect the transport of energy and lepton number in the supernova core implies that further attention should be devoted to improve our understanding of the behavior of neutrino mixing in dense media and especially of fast pairwise conversion. 
The physics of flavor mixing in the proximity of supernova shocks remains to be clarified~\cite{Friedland:2020ecy} as well as the role of turbulent fluctuations~\cite{Borriello:2013tha}. In addition to an improved understanding of neutrino physics, the self-consistent modeling of New Physics scenarios and an improved understanding of the impact of such scenarios on the core-collapse physics will be crucial to place robust constraints in the near future. One of the goals on the theoretical front would be an end-to-end consistent simulation of the burst physics and related detectable signatures. A consistent modeling of the neutrino signal in the neutrino telescopes and a good understanding of the detector background are also necessary. 

Within the multi-messenger scenario, neutrinos are one of  messengers. In order to maximize the amount of information that can be extracted from the other messengers,  it is important to pin down all the uncertainties on the expected neutrino signal. In particular, the determination of the angular location of the supernova burst in the sky and the employment of neutrinos for timing are also affected by existing uncertainties on flavor mixing~\cite{Sarfati:2021vym}.

As we approach the DSNB detection,  improved theoretical models and better understanding of the detector backgrounds are absolutely needed~\cite{Li:2022myd}. In order to maximally exploit the DSNB detection, it is  crucial to progress on the theoretical modeling of the supernova population and neutrino physics~\cite{Kresse:2020nto, Horiuchi:2020jnc, Moller:2018kpn,Lunardini:2012ne}; in particular, large theoretical uncertainties are  linked to the supernova rate and the role of binaries in the evolution of massive stars.

\subsubsection{Prospects for measurements beyond electron anti-neutrino}

Current detectors mainly measure core-collapse supernova neutrinos via the inverse beta decay channels, and are therefore mainly sensitive to electron anti-neutrinos. Detectors which can measure the neutral current channel will play an important role in making a measurement independent of oscillation effects, but will be low statistics. Next-generation dark matter detectors will view neutral current interactions with high enough statistics to extract oscillation-independent effects. 

DUNE will measure the electron neutrino channel with high statistics and low backgrounds. This will enable a measurement of the neutronization peak, a feature only visible in electron neutrinos, which carries information about neutrino ordering and can be used to calibrate the distance to the supernova. The THEIA detector, which is proposed to sit at the same site as DUNE, has the potential to act as a supernova trigger for DUNE thanks to the fast timing and lower threshold, which gives greater burst sensitivity, thus extending the DUNE detection horizon.

\subsubsection{Expanding the detection horizon for core-collapse supernovae}

The expected rate for a supernova to occur within our current detectable volume is only a few per century. The next big frontier in supernova neutrino detection will be to expand the detection horizon to include more galaxies. As the neutrino event rate scales with the inverse squared of the distance, this is a significant challenge. Hyper-Kamiokande will be large and essentially background-free, and so has good prospects to detect supernova bursts out to Andromeda. 

\subsubsection{Early warning for multi-messenger physics}

The SuperNova Early Warning System (SNEWS) is a global network of neutrino observatories formed in 2000 to report the detection of core-collapse supernovae across multiple detectors. Neutrinos arrive at Earth prior to the optical signal from the core-collapse explosion. Thus, neutrinos provide the possibility to alert observatories of an imminent supernova explosion even before it is visible in the sky; this is the challenge and purpose of the SNEWS network. It includes all sensitive neutrino detectors including Super-K (Japan), IceCube (South Pole), KamLAND (Japan), KM3NeT (Mediterranean), and HALO (Canada).  SNEWS2.0 is a follow-up effort that will play a crucial role in maximizing the potential for multi-messenger observations in the next core-collapse supernovae. SNEWS2.0 will improve the alert system by including more information about the supernova, such as the direction, as well as an alert for pre-supernova neutrino emission. Future neutrino detectors Hyper-K, JUNO, SNO+, NOvA, Baksan, Darwin, Darkside, DUNE, nEXO, XENONnT and LZ plan to connect to SNEWS2.0. 

\subsubsection{Measuring the diffuse supernova neutrino background}

The ongoing enrichment of the  Super-Kamiokande detector with Gadolinium and the upcoming JUNO detector provide very promising prospects for discovery ~\cite{Mirizzi:2015eza,Lunardini:2010ab,Beacom:2010kk}. Other proposed large-scale liquid scintillator or hybrid detectors, such as THEIA, would also be sensitive to the DSNB signal \cite{Theia:2022uyh}. In addition, the employment of coherent neutrino nucleus scattering detectors could provide  upper limits on the non-electron flavors~\cite{Suliga:2021hek}. Hyper-Kamiokande will move beyond discovery, towards precision measurements of the DSNB. This can allow for a measurement of the average neutrino energy, as well as the fraction of failed supernovae that end as black holes \cite{Hyper-Kamiokande:2018ofw}.  
\section{Atmospheric Neutrinos}
\subsection{Motivation and current understanding}
Atmospheric neutrinos are produced by collisions of primary cosmic rays with nuclei in the atmosphere, and by the decay of hadrons produced by these collisions.
The breakthrough discovery in atmospheric neutrinos was neutrino oscillation by Super-Kamiokande in 1998\cite{Fukuda:1998mi}.

Atmospheric neutrino detectors that are currently in operation and providing results for neutrino oscillation are mainly water (including ice) Cherenkov types such as Super-Kamiokande\cite{Abe:2017aap},
IceCube\cite{PhysRevLett.120.071801}, and ANTARES\cite{Albert2019MeasuringTA}.
These observations have led to precise measurements of neutrino oscillation parameters, especially for $\theta_{23}$ and $\Delta m^2_{32}$.
Recently, the precise observations and large statistics of Super-Kamiokande have made it possible to study all the sub-leading effects of neutrino oscillations.
Here, they show that the Earth-matter effect plays an important role in neutrino oscillations, which resolves mass ordering, two possible $\theta_{23}$ regions, and $\delta_{CP}$.
The results prefers normal mass ordering and 1st $\theta_{23}$ octant and $\delta_{CP}\sim 3/2\pi$ although not with large probability around one sigma level.

In addition to that, the significance of the appearance of tau neutrinos due to atmospheric neutrino oscillations was reported to be $\sim 5$ sigma in Super-Kamiokande\cite{Li:2017dbe} 
and 3.2 sigma in IceCube\cite{PhysRevD.99.032007}, and those results were consistent.
These experiments also reported the search for sterile neutrinos, although no significant signal was found\cite{Abe:2014gda, PhysRevD.95.112002, Albert2019MeasuringTA}.
Such a precise measurement of neutrino oscillation parameters and search for various phenomena caused by neutrino oscillation are tasks to be achieved in future atmospheric neutrino experiments.

Another topic includes the precise measurement of atmospheric neutrino fluxes.
In recent years, several results, both theoretical and observational, have been reported.
Several groups, such as HKKM\cite{PhysRevD.83.123001}, Bartol\cite{PhysRevD.70.023006}, and FLUKA group\cite{BATTISTONI2003269}, have made theoretical predictions.
The differences in flux prediction by these models comes from the choices of the hadron interaction model and the measurement of the primary cosmic-ray spectrum, and is about 10\%.
On the other hand, atmospheric neutrino flux measurements have been reported from several experiments, most recently from Super-Kamiokande\cite{PhysRevD.94.052001}, IceCube\cite{PhysRevD.83.012001, PhysRevD.91.122004, PhysRevLett.110.151105}, and ANTARES\cite{ANTARES2013}.
Their results are consistent with the predictions with uncertainties.
In addition, Super-Kamiokande also reports the directional dependence of the atmospheric neutrino flux on the rigidity cutoff and the long-term variability expected due to solar activity, which are also consistent with the predictions.
More accurate atmospheric neutrino flux predictions and observations are also future tasks.

One more thing to be pointed out is that the atmospheric neutrino events are background to nucleon decay searches (proton decay, neutron-antineutron oscillation, etc.), and cosmogenic dark matter searches, thus, understanding the atmospheric neutrino flux and interaction is important for those searches.

\subsection{Prospects for future measurements}

\subsubsection{Prediction of atmospheric neutrino flux}
Recently, two groups (Nagoya and Bartol) have repeated the HKKM and Bartol flux calculations.
Compared to the original version, the model is easier to tune because the primary flux and hadron production data have been greatly improved.
The model improvements are motivated by various on-going and future experiments.
In this calculation, the uncertainties of the hadron interaction is critical.
This theoretical group will consider the beam experiment whose kinematic region is important for atmospheric neutrino production.
Various beam experimental data already exist, but they are insufficient, therefore, it is important to obtain the results of the currently planned experiments, such as NA61/SHINE, EMPHATIC, protoDUNE-II and Forward Physics Facility at the LHC, as soon as possible.

\subsubsection{Neutrino oscillation measurements}
\paragraph{Super-Kamiokande and Hyper-Kamiokande}
As mentioned in the previous section, atmospheric neutrino measurements are sensitive to mass ordering, the two possible $\theta_{23}$ regions, and $\delta_{CP}$ due to the Earth-matter effect.
There are several challenges, such as flux uncertainties, neutrino/anti-neutrino separation, and reconstruction of multi-particle final states.
A new phase of gadolinium-doped Super-Kamiokande began in 2020, which should help more effective separation of neutrino/anti-neutrino.

In Hyper-Kamiokande\cite{HKDR2018}, the sensitivity of these parameters becomes quite good because of their huge statistics.
For example, it is expected in ten years observation to resolve the mass ordering at three sigma level for both normal and inverted assumptions and when $\sin^2\theta_{23}>0.53$, determine the $\theta_{23}$ octant at two sigma level when $|\theta_{23}-45|>4^{\circ}$, and make it possible to realize a precise $\delta_{CP}$ measurement when combined with the beam neutrino measurement.
For the appearance of tau neutrinos, more than 700 events are expected to be detected in 10 years of operation.
This will allow us to obtain a more accurate cross section measurement of charged-current $\nu_{\tau}$ interaction.
Hyper-Kamiokande is expected to improve the bounds on various non-standard neutrino properties obtained by Super-Kamiokande, such as the presence of sterile neutrinos, non-standard neutrino interactions with matter, and Lorentz invariance violation, by combining data from intermediate detectors several kilometers from the beam and atmospheric neutrino observations.

\paragraph{IceCube}
IceCube has the sensitivity to observe atmospheric neutrinos in a very wide energy range, from the order of GeV to 100~TeV.
The sensitivity over a wide energy range enables not only precise measurement of `conventional' atmospheric neutrinos, but also 'prompt' neutrinos from the decay of heavy hadrons which have not yet been discovered.
The IceCube Upgrade improves the resolution of neutrino energy and direction and lowers the energy threshold, allowing especially for improved neutrino oscillation sensitivity in atmospheric neutrinos.
In particular, the measurement of the matter effect of atmospheric neutrinos by IceCube Upgrade, combined with the JUNO experiment, determines the mass ordering with a sensitivity of 5 sigma after 3$\sim$7 years of observation\cite{PhysRevD.101.032006}.
In addition, the atmospheric neutrino high-energy region is well suited for searching for various BSM physics, such as sterile neutrino, Lorentz violations, non-standard interactions, and future IceCube experiment is expected to provide a lot of knowledge into these physics.

\paragraph{DUNE}
DUNE can measure atmospheric neutrinos well in the sub-GeV energy region.
This is due to its ability to efficiently tag protons.
The expected number of events in 400 kton-year is $\sim$5000 (for charged-current 0p0$\pi$), $\sim$9000 (for 1p0$\pi$), and $\sim$250 (for 2p0$\pi$).
This will lead, for example, a certain area containing $\delta_{CP}=\pi$ can be excluded at the three sigma level for an input value of $\delta_{CP}=3\pi/2$\cite{PhysRevLett.123.081801}.
DUNE will be also possible to provide a new window of BSM scenario from the atmospheric neutrino data.

\paragraph{Theia}
Theia is a proposed multi-kiloton hybrid optical neutrino detector that takes full advantage of scintillation and Cherenkov light information\cite{THEIA20}.
Various physics potentials have been evaluated, as well as the potential to utilize atmospheric neutrino observation data.
Especially, a placement of Theia in the LBNF beamline could further explore various non-standard neutrino properties mentioned in the discussion of Hyper-Kamiokande.

\section{High energy astrophysical neutrinos}

\subsection{Motivation and current understanding}

High energy astrophysical neutrinos emerge from Nature's accelerators: powerful astrophysical systems that allow particles to be accelerated to energies significantly higher than achievable with man made accelerators. These neutrinos can help us to understand the dense environments in which they are created, and give access to particle physics phenomena at high energies. 

The field of experimental high energy neutrino astrophysics has emerged only over the last decade. In the coming decade, we need to move beyond the discovery phase to robustly measure and characterize of sources. Moreover, we must further extend our energy reach in order to access source classes connected to the highest energy particles in our Universe. 

\subsubsection{Sources of high energy astrophysical neutrinos}

High energy neutrinos can be produced when accelerated protons interact with the surrounding environment such that charged pions are formed, which subsequently decay into neutrinos \cite{stecker1991high}. This can either happen through inelastic hadronuclear reactions (protons interacting with protons or neutrons, pp or pn), or through photohadronic interactions (protons interacting with photons, p$\gamma$) \cite{Anchordoqui:2013dnh}. 

Cosmic accelerators are good candidates for the production of high energy neutrinos. The dense environment has sufficient ambient material to act as a target and shocks can accelerate protons to high energies. The IceCube Neutrino Observatory made the first discovery of high energy astrophysical neutrinos in 2013 \cite{IceCube:2013low}. Since then, several neutrinos events have been detected in likely coincidence with cosmic accelerators \cite{IceCube:2018dnn, IceCube:2018cha, Stein:2020xhk, IceCube:2019cia,Reusch:2021ztx,Pitik:2021dyf}. However, more work remains to be done on the modeling of the production of high-energy particles within a multi-messenger framework that takes into account the population properties of these sources.

Other potential sources of high energy astrophysical neutrinos include those produced in dark matter annihilation \cite{Hiroshima:2017hmy, Arguelles:2019ouk} and the so-called GZK neutrinos expected from interactions between cosmic rays and the cosmic microwave background \cite{Greisen:1966jv, Zatsepin:1966jv}.

\subsubsection{Experimental landscape}

Current experimental efforts to measure and characterize the flux of high energy astrophysical neutrinos is dominated by neutrino telescopes, which use optical modules to detect the photons produced in neutrino interactions in ice or water. Neutrinos detected by these experiments range from GeV to PeV in energy. The IceCube Neutrino Observatory \cite{IceCube:2016zyt} is the largest operating detector, instrumenting a cubic kilometer of glacial ice at the geographic South Pole. Other neutrino telescopes in operation are located in the Northern hemisphere with the aim to measure neutrinos from the Southern sky. These include ANTARES \cite{ANTARES:2011} in the Mediterranean Sea and NT200+ \cite{BAIKAL:2006nib} in Lake Baikal. 

To observe the expected low flux of neutrinos with energies above a few PeV, new detection techniques are required. Detectors that utilize the Asakaryan effect, which produces coherent radio emission in ice and rock, are promising. Due to the long attenuation length of photons in this wavelength, sparse instrumentation can view the large volumes required to measure the anticipated low flux of these neutrinos. In-ice experiments such as ARIANNA \cite{ARIANNA:2019scz} and ARA \cite{Allison:2011wk} and balloon borne experiments such as ANITA \cite{ANITA:2008mzi} have served as prototypes of the technique. RNO-G \cite{RNO-G:2020rmc} is currently deploying instrumentation in Greenland and, when completed, will be the largest radio neutrino detector to date.

\subsubsection{High energy astrophysical neutrinos as probes for particle physics}

Neutrinos from astrophysical accelerators provide a unique opportunity to probe particle physics at the highest energies. Interactions of these neutrinos as they pass through the Earth allow us to measure the cross sections with neutrino energies up to $\sim$ a PeV, compared with below 400 GeV from neutrino beams \cite{IceCube:2017roe, Bustamante:2017xuy}. Measurements of the flavour ratio of astrophysical neutrinos has constrained flavour-changing processes, especially those predicted by new and beyond standard model physics \cite{Beacom:2003nh, Pakvasa:2007dc,Song:2020nfh}. The long propagation lengths of astrophysical neutrinos have also allowed constraints of energy-dependent tests for fundamental neutrino physics \cite{Ackermann:2019cxh}. Current indirect detection limits on the cross section for dark matter with masses above 1 TeV are driven by searches from ANTARES and IceCube \cite{Arguelles:2019ouk}.

In order to improve our current understanding, we require both higher statistics of astrophysical neutrinos with energies up to PeV as well as measurements from the yet-unexplored energy region above a few PeV. 

\subsubsection{High energy astrophysical neutrinos as probes for astrophysical environments}

The search for high energy neutrinos was born from another mystery in astrophysics: what are the sources of the high energy cosmic rays that bombard Earth's atmosphere? High energy neutrinos are inherently tied to hadronic activity and, unlike charged cosmic rays, neutrinos can travel directly from their point of creation to our detectors without interacting en route. Sources of high energy neutrinos should therefore be tied to the sources of cosmic rays. In 2018, IceCube announced for the first time that an astronomical source of these neutrinos could be identified when neutrinos from the direction of blazar TXS 0506 +056 coincided with an enhancement in gammas as detected in Fermi \cite{IceCube:2018dnn}. A subsequent follow up search from this direction showed enhanced neutrino emission in 2014 - 2015 that was uncorrelated with gamma emission \cite{IceCube:2018cha}. More recently, an IceCube neutrino alert has been linked to a tidal disruption event or a superluminous supernova \cite{Stein:2020xhk,Pitik:2021dyf}, and an analysis of ten years of IceCube data has indicated high energy neutrino emission from the nearby active galaxy NGC 1068 \cite{IceCube:2019cia}. 

The study of the properties of the high energy astrophysical flux also gives insight to the acceleration properties and material properties of the dense astrophysical environments in which they are produced \cite{Ackermann:2019ows}. In particular the flavour composition and the energy shape of the neutrino flux has been well studied using the astrophysical neutrinos detected in IceCube \cite{IceCube:2020wum}.

\subsection{Future Prospects}

\subsubsection{Future Experiments and technologies}

\subsubsubsection*{In-ice optical and radio}

\textbf{IceCube-Gen2} \cite{IceCube-Gen2:2020qha} is an envisioned next-generation neutrino telescope that would significantly advance the study of high energy astrophysical neutrinos. A planned optical extension with photosensors that have  $\sim$ 3 times the photo collection efficiency would cover 8 times more instrumented volume compared to current IceCube. The addition of a $\sim$ 500 km radio array allows for the energy reach to extend to the EeV scale. IceCube-Gen2 will make firmer associations between astrophysical neutrinos and their sources, will improve the characterization of the properties of the astrophysical neutrino flux, will better understand the propagation of high energy particles in the Universe, and will probe fundamental physics at high energies. A factor of ten more astrophysical neutrinos with improved directional reconstruction is anticipated, allowing IceCube-Gen2 to see sources that are five times fainter compared with current technology and allowing for fewer false coincidences with potential sources. The enhanced detector will provide more opportunities for multi-messenger detection, especially in the electromagnetic regime but also potentially in connection with gravitational waves \cite{IceCube-Gen2:2020qha}. The improved statistics will allow for characterization of the energy spectrum and flavor composition over a wide range of energies, which can be used to both probe the acceleration mechanism of cosmic rays as well as to search for new flavor-violating physics. The extension to neutrinos energies up to EeV in particular will allow for the study of neutrinos connected to sources of the highest energy cosmic rays and to probe the predicted GZK neutrinos. 

In the near future, Northern hemisphere telescopes on the scale of IceCube will come online, including \textbf{KM3NeT} \cite{KM3Net:2016zxf} in the Mediterranean Sea, \textbf{Baikal-GVD} \cite{Baikal-GVD:2019kwy} in Lake Baikal, and \textbf{P-ONE} in the Pacific Ocean near the coast of Canada \cite{P-ONE:2020ljt}. Together, these detectors will improve the discovery potential for high energy neutrinos from the Southern sky by three orders of magnitude \cite{Schumacher:2021hhm}. 

\subsubsubsection*{Surface detection of earth-skimming tau neutrinos}

As high energy tau neutrinos pass through matter, charged current interactions produce a tau particles. The subsequent decays create air showers that can be detected either optically, using  water Cherenkov detectors, or via geomagnetic radio emission using antennas.  

\textbf{TAMBO} (Tau Air-Shower Mountain-Based Observatory) \cite{Romero-Wolf:2020pzh} is a proposed detector to be built in Peru with the aim to measure Earth-skimming astrophysical tau neutrinos in the 100 TeV to 100 PeV energy range. Tau-neutrino-induced air showers are detected in cubic-meter-scale water Cherenkov detectors placed $\sim$ 100 m apart. With the nominal design of 22,000 tanks, it is expected that the TAMBO will have an effective area $\sim$ 10 times larger than the current IceCube effective area for tau neutrinos at $\sim$ 3 PeV [CITE HE and UHE whitepaper]. The southern hemisphere site will allow TAMBO to view the galactic centre and the strategic location within a valley allows a high geometric acceptance for air shower production. 

\textbf{Trinity} is a proposed system of 18 air-shower Cherenkov telescopes optimized for detecting Earth-skimming neutrinos with energies between 10 PeV and 1 EeV. The intended Trinity site is in the Northern hemisphere and so it will view the northern sky, making it complementary to the astrophysical neutrino measurements from IceCube. A prototype will be deployed on Frisco Peak, Utah in 2022.

\textbf{BEACON} (The Beamforming Elevated Array for COsmic Neutrinos) \cite{Wissel:2020sec} is a concept that uses radio antennas to measure the air showers from high energy astrophycial taus neutrinos. It is currently in the demonstration phase, with a prototype deployed at the White Mountain Research Station of California. The final design envisions antennas deployed at several sites around the world for a full sky coverage. Within 3 years of observations with a full-scale instrument consisting of 1000 stations, BEACON is expected to achieve a sensitivity comparable to pessimistic models of cosmogenic neutrinos.

\textbf{GRAND} (Giant Radio Array for Neutrino Detection) is a planned large-scale radio observatory that intends to view earth-skimming tau neutrinos with ultrahigh energies from 100 PeV to 100 EeV. The final envisioned configuration consists of $\sim$ 20 geographically separate and independent sub-arrays each consisting of 104 antennas that view a total of $\sim$ 10000 cubic kilometers. Due to the large size of the array, sub-degree angular resolution of the incoming neutrino direction is anticipated. This will enable searches for point sources of ultrahigh energy astrophysical neutrinos. A prototype of 35 antennas was deployed in 2018 in the Tain Shan mountains in China \cite{GRAND:2018iaj}. Plans for a larger prototype consisting of 300 antennas is currently underway \cite{GRAND_LOI}.

\subsubsubsection*{Balloon-borne}

\textbf{PUEO} (Payload for Ultrahigh Energy Observations) \cite{PUEO:2020bnn} is the proposed successor to the ANITA balloon experiment. PUEO will aim to measure astrophysical neutrinos with energies above 1 EeV and is expected to have more than an order of magnitude more sensitivity than ANITA below 30 EeV. Like ANITA, PUEO will view a large volume of Antarctic ice from a long-duration balloon. Expected improvements for PUEO include more instrumentation as well as an improved triggering system. 

\subsubsubsection*{Space missions}

\textbf{POEMMA} (Probe Of Extreme Multi-Messenger Astrophysics) is a NASA space-based mission designed to measure ultra high energy cosmic rays and cosmic neutrinos by viewing the optical signal produced in neutrino-induced air showers in the Earth's atmosphere. Two cameras will facilitate two complementary detection techniques: an ultraviolet camera that allows for observations of florescence and an optical camera for the detection of Cherenkov light. \textbf{EUSO-SPB2} (The Extreme Universe Space Observatory aboard a Super Pressure Balloon 2) will provide useful input to the detection strategy of POEMMA by demonstrating the detection technique and by measuring key backgrounds.  EUSO-SPB2 has been approved and will launch in Spring of 2023.

\subsubsubsection*{Radar echo technique}

The \textbf{RET} (radio echo telescope) \cite{RadarEchoTelescope:2021rca} is a proposed experiment exploring a new detection technique for neutrinos above a PeV in energy. RET uses the radar echo method, where the ionization deposits induced in dense material such as ice by high energy cascades are detected via the reflection of radio waves. RET recently demonstrated the detection concept using an electron beam at SLAC to mimic the ionization produced in the cascade. This method would bridge the energies visible with optical detectors and proposed UHE detectors such as radio arrays. 

\subsubsection{High energy astrophysical neutrinos as probes for particle physics}

\subsubsubsection*{Cross section measurements}

IceCube has measured the neutrino cross section for neutrino energies between 6.3 TeV and 980 TeV \cite{IceCube:2017roe} with an uncertainty of 30-40\%. The measurement agrees well with standard model predictions, but the enhanced statistics expected with the next generation of experiments will allow for precision tests to beyond standard model phenomena such as extra dimensions, leptoquarks or sphalerons, as well as probe down to the regime where non-perturbative QCD effects are expected to start becoming important. 

Future detectors will have the ability to explore effects related to W-boson production, including hidden Glashow resonances due to neutrino-nucleus collisions \cite{Alikhanov:2015kla} and processes affecting neutrino-photon interactions \cite{Seckel:1997kk}. 

\subsubsubsection*{Dark matter}

Weakly interacting massive particles (WIMPs) are primary candidates for the observed dark matter in our Universe. High energy astrophysical neutrinos can indirectly probe WIMPs if they decay to neutrinos. Future neutrino telescopes will probe new phase space of WIMP cross section and mass currently allowed.

\subsubsubsection*{New physics}

In general the long path lengths and large energy ranges provided by astrophysical neutrinos make them good candidates for searching for new physics phenomena. The higher energy reach of future detectors could provide a completely new regime in which to search for unexpected physics. For a more complete description, we refer the reader to \cite{Ackermann:2022rqc}. In particular, the search for neutrino decay, secret or non-standard neutrino interactions, BSM extensions to fundamental symmetries, quantum gravity and exotic scenarios such as magnetic monopoles can be probes with the next generation of neutrino telescopes . 

\subsubsection{High energy astrophysical neutrinos as probes for astrophysical environments}

\subsubsubsection*{Particle acceleration  in cosmic sources}

With the expected high statistics measurements from future telescopes, measurements of the detailed properties of accelerated particles are expected. This is especially interesting as the energies for particle acceleration as well as the plasma processes in the dense environments of these systems cannot be
probed directly by other means. Precision measurements of the spectral features will allow to study cosmic-ray interaction processes and maximum acceleration energies.  

\subsubsubsection*{Sources of astrophysical neutrinos}

There have been likely associations between high energy neutrinos and blazars, star burst galaxies, tidal disruption events, and supernovae. The next generation of neutrino telescopes will make high significance observations of the brightest sources and allow for characterization of the spectral properties of the neutrinos as well as the flaring behaviour of the emission, and the source properties \cite{Bustamante:2019sdb,Bustamante:2020bxp,Ando:2015bva,Tamborra:2015fzv, Guarini:2021gwh,Pitik:2021xhb,Fang:2021ylv,Fang:2020bkm,Fang:2017zjf}. 

Future detectors are also expected to resolve the question of the sources of cosmic rays emitted from our galaxy. IceCube has best sensitivity to the Northern sky, but near-term and future detectors in the Northern hemisphere as well as improved detectors such as IceCube-Gen2 can begin to probe high energy astrophysical neutrinos from the galactic centre. 

\section{Conclusion}
A rich landscape of physics can be pursued by utilising the broad range of neutrino natural sources, offering insights into the natural world, and the properties of these mysterious particles.  One aspect many of these areas have in common is that, as precision advances and we continue to push the bounds of our understanding, greater sensitivity is required, and it becomes ever more challenging to build a case for a dedicated detector.  Thus, we see a future in which synergies between physics topics play an ever increasing role, with emphasis moving towards large-scale detectors that offer a broad program of physics, across multiple sources and multiple energy scales.  This places considerable emphasis on novel technology development, as well as collaboration across traditional topical boundaries, in order to achieve a successful suite of next-generation experiments.
\section*{Acknowledgements}



The authors wish to thank all members of the NF04 Topical Group for their engagement and involvement in the discussions and workshops that led to this report, and  comments on the report itself.  In particular we wish to thank Maury Goodman, Georgia Karagiorgi, Spencer Klein, Lisa Koerner, and Michael Smy for their helpful input on the report.  We also wish to thank those who presented overview talks on many of these topics at dedicated and shared workshops as part of the Snowmass process.  We wish to thank all Convenors of Topical Groups under the Neutrino Frontier for invaluable discussions, and a most heartfelt thank you to Patrick Huber, Kate Scholberg, and Elizabeth Worcester for their stalwart guidance throughout this process.

\clearpage

\appendix
\section{Letters of Interest submitted to NF04}
The tables below list all LOIs submitted to NF04 as part of the Snowmass process.

\begin{table*}[!h]
\centering
\caption{LOIs submitted to NF04 part i}
\begin{tabular}{l} 
\hline\noalign{\smallskip}
\noalign{\smallskip}\hline\noalign{\smallskip}
CF/SNOWMASS21-CF1\_CF0-NF10\_NF4-IF3\_IF0\_Ethan\_Brown-034.pdf	\\
CF/SNOWMASS21-CF1\_CF0-NF10\_NF4-IF5\_IF4\_Vahsen-189.pdf	\\
CF/SNOWMASS21-CF1\_CF2-NF5\_NF4-IF8\_IF0-CompF0\_CompF0-UF2\_UF3\_Matthew\_Szydagis-236.pdf	\\
CF/SNOWMASS21-CF7\_CF0-NF10\_NF4-IF10\_IF4-139.pdf	\\
CF/SNOWMASS21-CF7\_CF0-NF10\_NF4-IF10\_IF4\_Wissel-088.pdf	\\
CF/SNOWMASS21-CF7\_CF0-NF10\_NF4\_S.\_Barwick-013.pdf	\\
CF/SNOWMASS21-CF7\_CF0-NF4\_NF0-IF10\_IF0-118.pdf	\\
CF/SNOWMASS21-CF7\_CF0-NF4\_NF0-IF10\_IF0\_Vieregg-222.pdf	\\
CF/SNOWMASS21-CF7\_CF0-NF4\_NF10-TF11\_TF0\_Tonia\_Venters-156.pdf	\\
CF/SNOWMASS21-CF7\_CF1-NF4\_NF0-216.pdf	\\
CF/SNOWMASS21-CF7\_CF1-NF4\_NF3-TF11\_TF0\_Mauricio\_Bustamante-048.pdf	\\
CF/SNOWMASS21-CF7\_CF3-NF4\_NF0\_Jaime\_Alvarez-Muniz-140.pdf	\\
CF/SNOWMASS21-CF7\_CF6-EF6\_EF0-NF4\_NF10-IF10\_IF0\_Joerg\_Hoerandel-117.pdf	\\
CF/SNOWMASS21-CF7\_CF6-NF4\_NF10\_Ignacio\_Taboada-092.pdf	\\
CF/SNOWMASS21-CF7\_CF6-NF4\_NF10\_Nepomuk\_Otte-202.pdf	\\
IF/SNOWMASS21-IF6\_IF8-NF4\_NF9-CF1\_CF2\_Rick\_Gaitskell-172.pdf	\\
NF/SNOWMASS21-NF10\_NF4-CF1\_CF0-IF8\_IF0-UF1\_UF3-137.pdf	\\
NF/SNOWMASS21-NF10\_NF4-CF7\_CF0-IF10\_IF0\_Prohira-109.pdf	\\
NF/SNOWMASS21-NF10\_NF4-CF7\_CF1-TF11\_TF0\_Darren\_Grant-105.pdf	\\
NF/SNOWMASS21-NF1\_NF4-RF4\_RF5\_Aurisano-154.pdf	\\
NF/SNOWMASS21-NF1\_NF4\_Pedro\_Ochoa-034.pdf	\\
NF/SNOWMASS21-NF1\_NF4\_SNOplus-185.pdf	\\
NF/SNOWMASS21-NF2\_NF4\_Sousa\_Thakore-134.pdf	\\
NF/SNOWMASS21-NF3\_NF4-CF7\_CF0-TF11\_TF0\_Segev\_BenZvi-043.pdf	\\
NF/SNOWMASS21-NF3\_NF4\_Russell\_Neilson-017.pdf	\\
NF/SNOWMASS21-NF4\_NF0\_Matthew\_Szydagis-163.pdf	\\
NF/SNOWMASS21-NF4\_NF1-RF4\_RF0-CF7\_CF1\_SUPERK-050.pdf	\\
NF/SNOWMASS21-NF4\_NF10-CF1\_CF0\_Alec\_Habig-124.pdf	\\
NF/SNOWMASS21-NF4\_NF10-CF1\_CF7-020.pdf	\\
NF/SNOWMASS21-NF4\_NF10-CF7\_CF0-TF11\_TF0-IF10\_IF0\_Mauricio\_Bustamante-195.pdf	\\
NF/SNOWMASS21-NF4\_NF10-CF7\_CF1\_Mauricio\_Bustamante-044.pdf	\\
NF/SNOWMASS21-NF4\_NF10-IF6\_IF0\_Orebi\_Gann-089.pdf	\\
NF/SNOWMASS21-NF4\_NF10\_OceanBottomDetector-201.pdf	\\
NF/SNOWMASS21-NF4\_NF10\_PTOLEMY-021.pdf	\\
NF/SNOWMASS21-NF4\_NF10\_Pattavina-076.pdf	\\
NF/SNOWMASS21-NF4\_NF10\_Steve\_Biller-013.pdf	\\
NF/SNOWMASS21-NF4\_NF10\_Steve\_Biller-060.pdf	\\
\noalign{\smallskip}\hline
\end{tabular}
\end{table*}

\begin{table*}[!h]
\centering
\caption{LOIs submitted to NF04 part ii}
\begin{tabular}{l} 
\hline\noalign{\smallskip}
\noalign{\smallskip}\hline\noalign{\smallskip}
NF/SNOWMASS21-NF4\_NF3-CF1\_CF7-TF9\_TF11-039.pdf	\\
NF/SNOWMASS21-NF4\_NF3-CF7\_CF6-TF11\_TF9\_DUNE-054.pdf	\\
NF/SNOWMASS21-NF4\_NF3\_Matthew\_Strait-090.pdf	\\
NF/SNOWMASS21-NF4\_NF5-TF11\_TF0-158.pdf	\\
NF/SNOWMASS21-NF4\_NF6-CF7\_CF3-TF9\_TF11-IF2\_IF10\_Wissel-064.pdf	\\
NF/SNOWMASS21-NF5\_NF4-084.pdf	\\
NF/SNOWMASS21-NF5\_NF4-RF4\_RF0\_Savarese-127.pdf	\\
NF/SNOWMASS21-NF5\_NF4\_Baudis-085.pdf	\\
NF/SNOWMASS21-NF6\_NF4-IF2\_IF8-139.pdf	\\
NF/SNOWMASS21-NF8\_NF4-CF1\_CF7\_Evan\_Grohs-174.pdf	\\
NF/SNOWMASS21-NF8\_NF4-CF3\_CF7-TF9\_TF8\_Johns-121.pdf	\\
NF/SNOWMASS21-NF9\_NF4\_Zimmerman,Eric-069.pdf	\\
TF/SNOWMASS21-TF11\_TF9-NF4\_NF10-IF6\_IF0\_Alex\_Friedland-085.pdf	\\
\noalign{\smallskip}\hline
\end{tabular}
\end{table*}

\newpage
\section{White Papers submitted to NF04}
The following lists all White Papers submitted directly to NF04 as part of the Snowmass process.

\begin{itemize}
    \item J. Aalbers, K. Abe, V. Aerne), F. Agostini, S. Ahmed Maouloud, D.S. Akerib, et al. ”A Next-Generation Liquid Xenon Observatory for Dark Matter and Neutrino Physics”, arXiv:2203.02309 [physics.ins-det] (pdf). (also under RF03, CF01, IF08, UF03)

\item Junhui Liao, Yuanning Gao, Zhuo Liang, Zebang Ouyang, Chaohua Peng, Fengshou Zhang, Lei Zhang, Jian Zheng, Jiangfeng Zhou. ”Introduction to a low-mass dark matter project, ALETHEIA: A Liquid hElium Time projection cHambEr In dArk matter“, arXiv:2203.07901 [astro-ph.IM] (pdf). (also under CF01, RF03, IF08, UF03)

\item Markus Ackermann, Sanjib K. Agarwalla, Jaime Alvarez-Muñiz, Carlos A. Argüelles, et al. ”High-Energy and Ultra-High-Energy Neutrinos”, arXiv:2203.08096 [hep-ph] (pdf). (also under CF07, TF09, IF10)

\end{itemize}

The following lists White Papers submitted to the Neutrino Frontier with interest to multiple topical groups:

\begin{itemize}
    \item Luis A. Anchordoqui, A. Ariga, T. Ariga, et al. “The Forward Physics Facility: Sites, Experiments, and Physics Potential”, arXiv:2109.10905 [hep-ph] (pdf). (also under EF0, RF0, CF0, TF0)
\item M. Askins, Z. Bagdasarian, N. Barros, E. W. Beier, et al. “Theia: Summary of physics program”, arXiv:2202.12839 [hep-ex] (pdf). (also under IF0, UF0)
\item Jonathan L. Feng, Felix Kling, Mary Hall Reno, Juan Rojo, Dennis Soldin, et al. ”The Forward Physics Facility at the High-Luminosity LHC“, arXiv:2203.05090 [hep-ex] (pdf). (also under EF0, RF0, CF0, TF0, AF0, IF0)
\item A. Abed Abud, B. Abi, R. Acciarri, et al. (DUNE Collaboration). ”DUNE Physics Summary”, arXiv:2203.06100 [hep-ex] (pdf). (also under RF04, CF01)
\item John Arrington, Joshua Barrow, Brian Batell, Robert Bernstein, Nikita Blinov, et al. ”Physics Opportunities for the Fermilab Booster Replacement”, arXiv:2203.03925 [hep-ph] (pdf). (also under RF0, TF0, AF0)
\item G. Bernardi, E. Brost, D. Denisov, G. Landsberg, M. Aleksa, D. d'Enterria, P. Janot, M.L. Mangano, et al. “The Future Circular Collider: a Summary for the US 2021 Snowmass Process”, arXiv:2203.06520 [hep-ex] (pdf). (also under EF0, RF0, TF0, AF0, IF0)
\item A. Abed Abud, B. Abi, R. Acciarri, M. A. Acero, M. R. Adames, G. Adamov, et al. ”A Gaseous Argon-Based Near Detector to Enhance the Physics Capabilities of DUNE“, arXiv:2203.06281 [hep-ex] (pdf). (also under IF0)
\item Dave Casper, Maria Elena Monzani, Benjamin Nachman, Costas Andreopoulos, Stephen Bailey, Deborah Bard, et al. ”Software and Computing for Small HEP Experiments”, arXiv:2203.07645 [hep-ex] (pdf). (also under EF0, RF0, CF0, CompF0)
\item Sunanda Banerjee, D. N. Brown, David N. Brown, Paolo Calafiura, Jacob Calcutt, Philippe Canal, et al. ”Detector and Beamline Simulation for Next-Generation High Energy Physics Experiments”, arXiv:2203.07614 [hep-ex] (pdf). (also under EF0, RF0, CF0, IF0, CompF02)
\item Jaret Heise. ”The Sanford Underground Research Facility”, arXiv:2203.08293 [hep-ex] (pdf). (also under RF0, CF0, IF0, CompF0, UF0, CommF0)
\item A. Alekou, E. Baussan, N. Blaskovic Kraljevic, M. Blennow, M. Bogomilov, E. Bouquerel, et al. ”The European Spallation Source neutrino Super Beam“, arXiv:2203.08803 [physics.acc-ph] (pdf). (also under RF0, CF0, AF0, UF0)
\item J. R. Alonso, J. M. Conrad, Y. D. Kim, S. H. Seo, M. H. Shaevitz, J. Spitz, D. Winklehner. ”IsoDAR Overview”, arXiv:2203.08804 [hep-ex] (pdf). (also under AF0, CommF01)
\item Emanuela Barzi, S. James Gates Jr., Roxanne Springer. ”In Search of Excellence and Equity in Physics“, arXiv:2203.10393 [physics.soc-ph] (pdf). (also under EF0, RF0, CF0, TF0, AF01, IF0, CompF0, UF0, CommF0)
\item Henning O. Back, Walter Bonivento, Mark Boulay, Eric Church, Steven R. Elliott, et al. ”A Facility for Low-Radioactivity Underground Argon”, arXiv:2203.09734 [physics.ins-det] (pdf). (also under CF0, IF09, UF04)
\item J. M. Campbell, M. Diefenthaler, T. J. Hobbs, S. Höche, J. Isaacson, F. Kling, et al. ”Event Generators for High-Energy Physics Experiments”, arXiv:2203.11110 [hep-ph] (pdf). (also under EF0, CF0, TF0, CompF02)
\item M. Endo, K. Hamaguchi, M. Ibe, T. Ishibashi, A. Ishikawa, M. Ishino, M. Ishitsuka, S. Kanemura, M. Kuriki, T. Mori, S. Moriyama, H. Nanjo, W. Ootani, and Y. Sato. ”Japan’s Strategy for Future Projects in High Energy Physics“, arXiv:2203.13979 [hep-ex] (pdf). (also under EF0, RF0, CF0, AF0, UF0)
\item E. Barzi, B. Barish, W. A. Barletta, I. F. Ginzburg, S. Di Mitri. ”High Energy \& High Luminosity $\gamma$$\gamma$ Colliders”, arXiv:2203.08353 [physics.acc-ph] (pdf). (also under EF0, RF0, CF0, TF0, AF03)
\item Sergey Pereverzev, Gianpaolo Carosi, Viacheslav Li. ”Superconducting Nanowire Single-Photon Detectors and effect of accumulation and unsteady releases of excess energy in materials”, arXiv:2204.01919 [quant-ph] (pdf). (also under CF0, IF06)

\end{itemize}


\renewcommand{\refname}{References}

\printglossary

\bibliographystyle{utphys}

\bibliography{references}

\end{document}